\documentclass[aps,prd,nofootinbib,twocolumn,superscriptaddress,preprintnumbers,balancelastpage,longbibliography]{revtex4-1}

\usepackage{aas_macros}
\usepackage[utf8]{inputenc}
\usepackage{amsmath,amssymb,mathtools,bm}
\usepackage{graphicx, color, hepunits}
\usepackage[dvipsnames]{xcolor}
\usepackage{float}
\usepackage{multirow}
 \usepackage{hyperref} 
\hypersetup{
    colorlinks=true,       
    linkcolor=blue,        
    citecolor=blue,        
    filecolor=magenta,     
    urlcolor=blue          
}
\usepackage[utf8]{inputenc}
\usepackage[english]{babel}
\usepackage{lipsum}

\usepackage{tensor}
\usepackage{ bbold }
\usepackage{tabularx}

\usepackage{dsfont}

\newcolumntype{Y}{>{\centering\arraybackslash}X}

\newcommand{\gagg}{g_{a \gamma \gamma}}

\newcommand{\es}[2] {\begin{equation} \label{#1} \begin{split} #2 \end{split} \end{equation}}

\begin{document}

\title{Upper limit on the axion-photon coupling from magnetic white dwarf polarization}

\author{Christopher Dessert}
\affiliation{Berkeley Center for Theoretical Physics, University of California, Berkeley, CA 94720, U.S.A.}
\affiliation{Theoretical Physics Group, Lawrence Berkeley National Laboratory, Berkeley, CA 94720, U.S.A.}
\affiliation{Leinweber Center for Theoretical Physics, Department of Physics, University of Michigan, Ann Arbor, MI 48109 U.S.A.}

\author{David Dunsky}
\affiliation{Berkeley Center for Theoretical Physics, University of California, Berkeley, CA 94720, U.S.A.}
\affiliation{Theoretical Physics Group, Lawrence Berkeley National Laboratory, Berkeley, CA 94720, U.S.A.}

\author{Benjamin R. Safdi}
\affiliation{Berkeley Center for Theoretical Physics, University of California, Berkeley, CA 94720, U.S.A.}
\affiliation{Theoretical Physics Group, Lawrence Berkeley National Laboratory, Berkeley, CA 94720, U.S.A.}

\date{\today}

\begin{abstract}
      
      Polarization measurements of thermal radiation from magnetic white dwarf (MWD) stars have been proposed as a probe of axion-photon mixing.  The radiation leaving the surface of the MWD is unpolarized, but if low-mass axions exist then photons polarized parallel to the direction of the MWD's magnetic field may convert into axions, which induces a linear polarization dependent on the strength of the axion-photon coupling $g_{a\gamma\gamma}$.  We model this process by using the formalism of axion-photon mixing in the presence of strong-field vacuum birefringence to show that of all stellar types MWDs are the most promising targets for axion-induced polarization searches.  We then consider linear polarization data from multiple MWDs, including SDSS J135141 and Grw+70$^\circ$8247, to show that after rigorously accounting for astrophysical uncertainties the axion-photon coupling is constrained to $|g_{a\gamma\gamma}| \lesssim 5.4 \times 10^{-12}$ GeV$^{-1}$ at 95\% confidence for axion masses $m_a \lesssim 3 \times 10^{-7}$ eV.  This upper limit puts in tension the previously-suggested explanation of the anomalous transparency of the Universe to TeV gamma-rays in terms of axions.  We identify MWD targets for which future data and modeling efforts could further improve the sensitivity to axions.

\end{abstract}
\maketitle

\section{Introduction}

Ultralight axion-like particles are hypothetical extensions of the Standard Model that could be remnants of new physics at energies well above those that may be probed by collider experiments~\cite{Ringwald:2014vqa,Choi:2020rgn,DiLuzio:2020wdo}.
For example, in String Theory compactifications it is common to find a spectrum of ultralight axions~\cite{Svrcek:2006yi,Arvanitaki:2009fg}.
At low energies the axions interact with the Standard Model through dimension-5 operators suppressed by the high scale $f_a \gtrsim 10^7$ GeV~\cite{Graham:2015ouw}.  In particular, an axion $a$ may interact with electromagnetism through the Lagrangian term ${\mathcal L} = g_{a\gamma\gamma} a {\bf E} \cdot {\bf B}$, where ${\bf E}$ and ${\bf B}$ are the electric and magnetic fields, respectively, and $g_{a \gamma\gamma} \propto 1/f_a$ is the  coupling constant.  In this work, we set some of the strongest constraints to-date on $g_{a\gamma\gamma}$ for low-mass axions using white dwarf (WD) polarization measurements.

\begin{figure}[!ht]
\begin{center}
\includegraphics[width=0.48\textwidth]{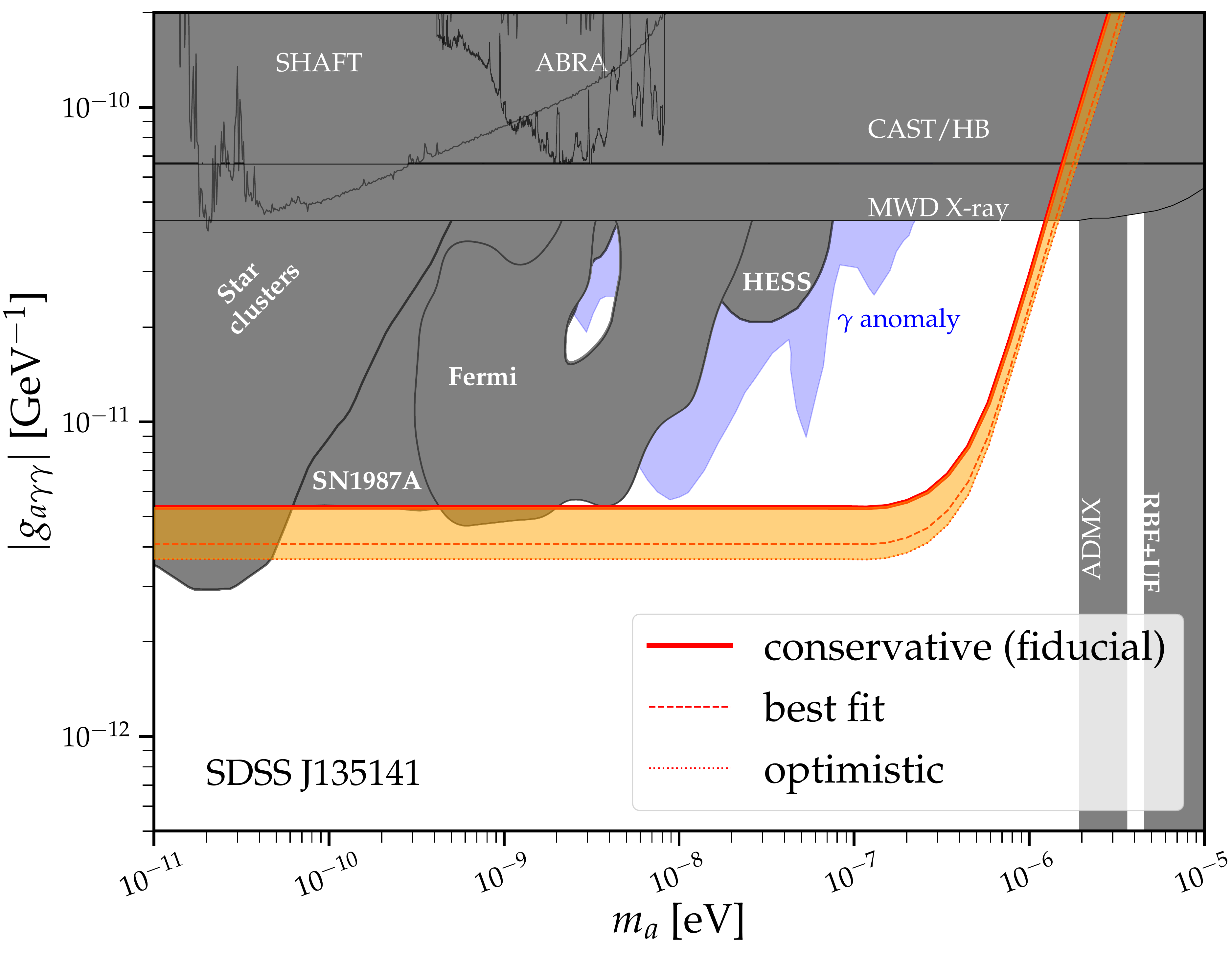}
\caption{\label{fig:J135_limits} 
Constraints on the axion-photon coupling $g_{a\gamma\gamma}$ arise from searches for axion-induced $X$-rays from super star clusters~\cite{Dessert:2020lil} and a nearby MWD~\cite{Dessert:2021bkv} in addition to gamma-rays from SN1987A~\cite{Payez:2014xsa}, searches for spectral irregularities with
Fermi-LAT~\cite{Fermi-LAT:2016nkz,Zhang:2018wpc} and H.E.S.S.~\cite{HESS:2013udx}, the CAST axion helioscope~\cite{Anastassopoulos:2017ftl}, HB star cooling~\cite{Ayala:2014pea}, and constraints from SHAFT~\cite{Gramolin:2020ict}, ABRACADABRA~\cite{Ouellet:2018beu,Salemi:2021gck}, ADMX~\cite{ADMX:2018gho,ADMX:2019uok}, and RBF+UF~\cite{DePanfilis:1987dk,Hagmann:1990tj} that are contingent on the axion being dark matter.  The fiducial 95\% upper limit from this work from the non-observation of linear polarization from SDSS J135141 is computed assuming the most conservative (at 1$\sigma$) magnetic field strength, MWD radius, and orientation. The shaded orange region shows how the limits change when considering astrophysical uncertainties; the dominant uncertainty is the inclination angle.  The limit found using the best-fit astrophysical parameters for the MWD is also indicated.
\vspace{-0.25in}
}
\end{center}
\end{figure}

Axions are notoriously difficult to probe experimentally due to their feeble interactions with the Standard Model.
The most powerful approach at present to probe ultralight axions purely in the laboratory is that employed by light shining through walls experiments, which leverage the fact that photons and axions mix in the presence of strong magnetic fields; the ALPS~\cite{Ehret:2010mh} experiment has constrained $|g_{a\gamma\gamma}| \lesssim 5 \times 10^{-8}$ GeV$^{-1}$ at 95\% confidence for axion masses $m_a \lesssim {\rm few} \times 10^{-4}$ eV.  The upcoming experiment ALPS-II~\cite{Bahre:2013ywa} may reach sensitivity to $|g_{a\gamma\gamma}| \lesssim 2 \times 10^{-11}$ GeV$^{-1}$ for a comparable mass range.  Going to lower coupling values, however, requires making use of astrophysical axion sources in order to access strong magnetic fields, longer distances, and higher luminosities.  For example, the CAST~\cite{Anastassopoulos:2017ftl} experiment (see Fig.~\ref{fig:J135_limits}) has set strong constraints on $g_{a \gamma\gamma}$ by looking for axions produced in the Sun and then converting to $X$-rays in the magnetic field of their detector, and the followup project IAXO~\cite{Armengaud:2014gea}  may be able to cover significant unexplored parameter space ($|g_{a\gamma\gamma}| \lesssim 4 \times 10^{-12}$ GeV$^{-1}$ for $m_a \lesssim 5 \times 10^{-3}$ eV).  
Purely astrophysical probes currently set the strongest constraints on $g_{a\gamma\gamma}$ at ultra-low axion masses.  Observations of horizontal branch (HB) star cooling~\cite{Ayala:2014pea} constrain $g_{a\gamma\gamma}$ at a level comparable to CAST ($|g_{a\gamma\gamma}| \lesssim 6.6 \times 10^{-11}$ GeV$^{-1}$, as illustrated in Fig.~\ref{fig:J135_limits}, for axion masses less than the keV scale).  The non-observation of gamma-rays from SN1987A --- which would be produced from Primakoff production in the supernova core and converted to photons in the Galactic magnetic fields --- leads to the limit $|g_{a\gamma\gamma}| \lesssim 5.3 \times 10^{-12}$ GeV$^{-1}$ for $m_a \lesssim 4.4 \times 10^{-10}$ eV~\cite{Payez:2014xsa} (but see~\cite{Bar:2019ifz}).  The non-observation of $X$-rays from super star clusters, which may arise from axion production in the stellar cores and conversion in Galactic magnetic fields, leads to the limit $|g_{a\gamma\gamma}| \lesssim 3.6 \times 10^{-12}$ GeV$^{-1}$ for $m_a \lesssim 5 \times 10^{-11}$ eV~\cite{Dessert:2020lil}.  Ref.~\cite{Reynolds:2019uqt} claims to constrain $|g_{a\gamma\gamma}| \lesssim 8 \times 10^{-13}$ GeV$^{-1}$ for $m_a \lesssim 10^{-12}$ eV using searches for $X$-ray spectral irregularities from the active galactic nucleus NGC 1275, though the magnetic field models in that work, and thus the resulting limits, are subject to debate~\cite{Libanov:2019fzq,Matthews:2022gqi}.

There are a number of astrophysical anomalies that favor axions at $|g_{a\gamma\gamma}|$ below current constraints.  For example, the unexplained transparency of the Universe to TeV gamma-rays may be explained by the existence of axions with $g_{a\gamma\gamma} \sim 10^{-12} - 10^{-10}$ GeV$^{-1}$ and $m_a \sim 10^{-9} - 10^{-8}$ eV (see Fig.~\ref{fig:J135_limits})~\cite{2011JCAP...11..020D,Essey:2011wv,Horns:2012fx,Meyer:2013pny,Rubtsov:2014uga,Kohri:2017ljt} (but see~\cite{Biteau:2015xpa,Dominguez:2015ama}).  The high-energy gamma-rays would convert to axions in the magnetic fields surrounding the active galactic nuclei sources and then reconvert to photons closer to Earth in the inter-galactic magnetic fields, effectively reducing the attenuation of gamma-rays caused by pair-production off of the extragalactic background light.  The gamma-ray transparency anomalies are constrained in-part by searches for spectral irregularities from gamma-ray sources with the H.E.S.S~\cite{HESS:2013udx} and Fermi-LAT~\cite{Fermi-LAT:2016nkz,Zhang:2018wpc} telescopes (but see~\cite{Libanov:2019fzq}).

Magnetic WDs (MWDs) are natural targets for axion searches because of their large magnetic field strengths, which can reach up to $\sim$$10^9$ G at the surface.  Ref.~\cite{Dessert:2021bkv} recently constrained the coupling combination $|g_{a\gamma\gamma} g_{aee}|$, with $g_{aee}$ the axion-electron coupling, using a {\it Chandra} $X$-ray observation of the MWD RE J0317-853.  Axions would be produced from electron bremsstrahlung within the MWD cores and then converted to $X$-rays in the magnetosphere.  Depending on the relation between $g_{aee}$ and $g_{a\gamma\gamma}$ the constraint on $g_{a\gamma\gamma}$ alone could vary from $|g_{a\gamma\gamma}| \lesssim {\rm few} \times 10^{-13}$ GeV$^{-1}$ to $|g_{a\gamma\gamma}| \lesssim 4.4 \times 10^{-11}$ GeV$^{-1}$ for $m_a \lesssim 5 \times 10^{-6}$ eV;  the most conservative constraint from that work is illustrated in Fig.~\ref{fig:J135_limits}. (See~\cite{Fortin:2018ehg,Fortin:2018aom,Buschmann:2019pfp,Fortin:2021sst} for similar searches using neutron stars (NSs) as targets.)  Note that WD cooling provides one of the most sensitive probes of the axion-electron coupling alone, since the axions produced by bremsstrahlung within the stellar cores provide an additional pathway for the WDs to cool~\cite{Raffelt:1985nj}.  

\begin{figure}[htb]
\begin{center}
\includegraphics[width=0.48\textwidth]{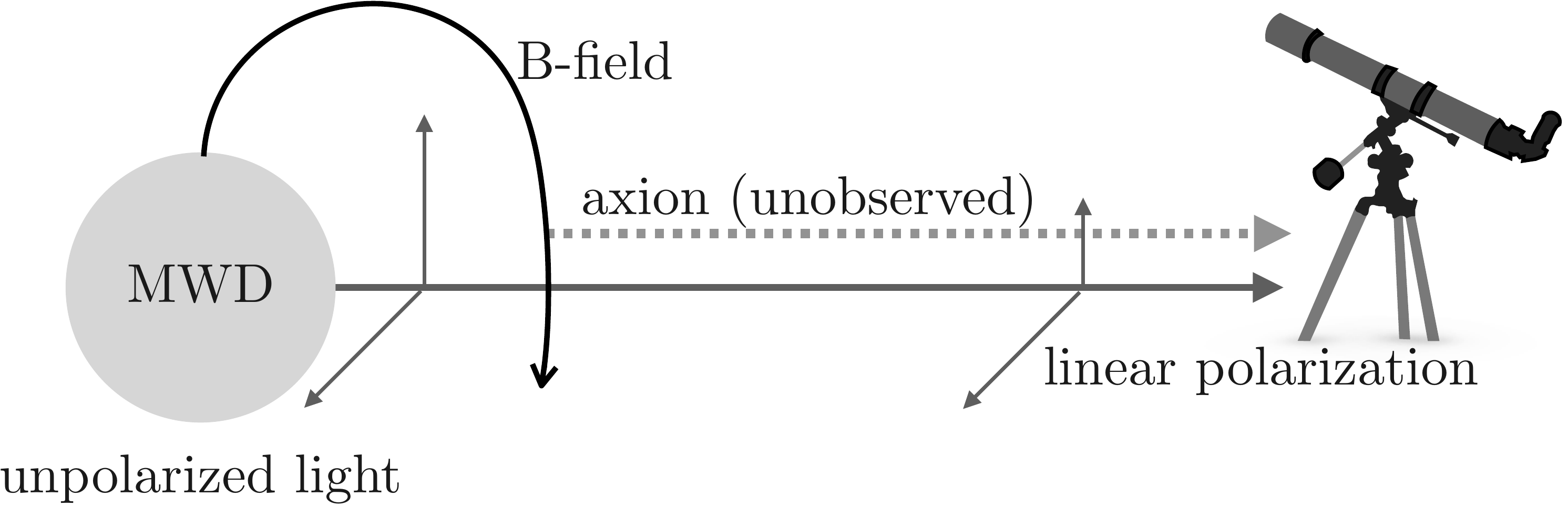} 
\caption{
The MWD emits thermal, unpolarized light, but this light may acquire a linear polarization when traversing the magnetosphere by photon-to-axion conversion.  Photons polarized along the direction of the transverse magnetic field may convert to axions, while those polarized in the orthogonal direction are unaffected.  Note that the conversion process may take place well away from the MWD surface.     
\label{fig:ill} }
\end{center}
\end{figure}

Refs.~\cite{Lai:2006af,Gill_2011} were the first to propose using MWD polarization measurements to constrain $g_{a\gamma\gamma}$.  The basic idea behind this proposal, which is the central focus of this work, is illustrated in Fig.~\ref{fig:ill}.
The MWD radiates thermally at its surface temperature.  The thermal radiation is unpolarized, but it may effectively acquire a linear polarization when traversing the magnetosphere because photons polarized parallel to the transverse magnetic fields may convert to axions, which are unobserved, while the orthogonal polarization direction is unaffected.  Ref.~\cite{Gill_2011} claimed that MWD linear polarization measurements of the MWDs PG 1031+234 and Sloan Digital Sky Survey (SDSS) J234605+38533 may be used to constrain $|g_{a\gamma\gamma}| \lesssim (5-9) \times 10^{-13}$ GeV$^{-1}$ for $m_a \lesssim {\rm few} \times 10^{-7}$ eV.  Here we critically reassess the upper limits from these MWDs and show that, while strong, the upper limits on $g_{a\gamma\gamma}$ from these MWDs are around an order of magnitude weaker than claimed in~\cite{Gill_2011}, when accounting for astrophysical uncertainties on the magnetic field and its geometry. Additionally, we identify two other MWDs --- SDSS J135141.13+541947.4 (hereafter SDSS J135141) and Grw 70$^\circ$8247 --- whose linear polarization measurements lead to strong constraints on $g_{a\gamma\gamma}$.  The upper limits on $g_{a\gamma\gamma}$ from this work represent the strongest to-date for ${\rm few} \times 10^{-9} \, \, {\rm eV} \lesssim m_a \lesssim 10^{-6}$ eV.  We show that the axion-induced polarization signal is determined only by the magnetic field strength and geometry far away from the MWD surface, outside of the atmosphere, where the free-electron plasma does not play an important role.  Lastly, we identify future MWD targets whose polarization observations could further constrain $g_{a\gamma\gamma}$ or lead to evidence for axions at currently un-probed coupling strengths.  We begin, in Sec.~\ref{sec:formalism}, by outlining the formalism for how to compute the axion-induced polarization signal.

\section{Axion-Induced Polarization}
\label{sec:formalism}

In this section we outline the formalism for computing polarization signals from astrophysical sources due to axion-photon mixing.  While we ultimately focus on MWDs in this work, we begin with a more general survey of possible astrophysical targets.  The basic idea behind this work is to focus on sources where the initial electromagnetic emission is known to be unpolarized but where the radiation must traverse regions of large magnetic field strengths before reaching Earth. Since photons polarized along the directions of the transverse magnetic fields may convert to axions, the presence of axions in the spectrum of nature will effectively induce a level of linear polarization whose degree depends on the strength of the axion-photon coupling.  
This process is illustrated for MWDs in Fig.~\ref{fig:ill}, where the relevant magnetic field is that directly surrounding the MWD.

The idea of searching for axion-induced polarization signals has been discussed in three main contexts: MWDs~\cite{Lai:2006af,Gill_2011}, NSs~\cite{Lai:2006af,Perna:2012wn}, and quasars~\cite{Jain:2002vx,Payez:2009vi,Agarwal:2009ic,Payez:2011sh,Agarwal:2011uk,Payez:2012vf,Masaki:2017aea,Galanti:2022iwb}.  In the first two cases the star is the source of both the initially-unpolarized photons and the strong magnetic fields. In the latter case, the magnetic fields are much weaker but they act over larger distances.  In this section we focus on polarization signals of the former type, where the star provides both the source of photons and magnetic fields, but first we briefly discuss the results of the quasar searches.  Ref.~\cite{Payez:2012vf} claims to constrain $|g_{a\gamma\gamma}| \lesssim {\rm few} \times 10^{-13}$ GeV$^{-1}$ for $m_a \lesssim {\rm few} \times 10^{-14}$ eV in order to not overproduce the measured optical polarization signals from distant quasars; this upper limit would be the most stringent to-date on low mass axions.  However, the results in~\cite{Payez:2012vf} are dependent on the strength of the assumed magnetic fields and plasma density profiles over distances $\sim$20 Mpc away from the sources.  Ref.~\cite{Payez:2012vf} assumed supercluster magnetic fields $\sim$2$\mu$G in strength and coherent over $\sim$100 kpc distances within 20 Mpc of the quasars.  On the other hand, simulations of supercluster magnetic fields~\cite{Dolag:2004kp,Dolag:2010ni,Marinacci:2017wew,Vazza:2017qge,Garcia:2020kxm} find that the fields are filamentary and typically orders of magnitude smaller than those assumed in~\cite{Payez:2012vf} at such large distances away from the clusters.  The field strengths increase in the clusters themselves, but so too does the free-electron density, which suppresses photon-to-axion conversion.  At present it seems likely that our knowledge of the supercluster-scale magnetic fields and plasma density profiles are not robust enough to claim a bound on $g_{a\gamma\gamma}$, which is why we focus on stellar sources for which the magnetic field profiles may be measured more precisely using {\it e.g.} the Zeeman effect and for which, as we will show, knowledge of the free-electron density is not necessary.  

\subsection{Analytic aspects of axion-induced polarization}

Consider an unpolarized monochromatic  beam of photons with frequency $\omega$ propagating through a medium with magnetic field profile ${\bf B}(s)$ and plasma-frequency profile $\omega_{\rm pl}(s)$, with $s$ the distance along the propagation direction.  The plasma frequency is sourced by 
free electrons for our purposes.  We will track the Stokes parameters, which in terms of the complex electric field ${\bf E}$ are defined by 
\es{eq:stokes}{
I &= | E_1 |^2 + |E_2|^2 \,, \qquad Q = |E_1|^2 - |E_2|^2 \\
U &= 2 {\rm Re}\left(E_1 E_2* \right) \,, \qquad V = -2 {\rm Im}\left(E_1 E_2* \right) \,,
}
with ${\bf x}_1-{\bf x}_2$ the transverse directions to the propagation direction ${\bf x}_3$.  The linear polarization fraction is conventionally defined by
\es{eq:linear}{
L_p \equiv {\sqrt{Q^2 + U^2} \over I} \,,
}
while the circular polarization fraction, which we will discuss less in this work, is $C_p \equiv V/I$.  The linear polarization is also specified by an angle in the ${\bf x}_1-{\bf x}_2$ plane $\chi$, with $\tan 2 \chi = U/Q$.  Note that we are interested in time-averaged quantities.  Thus, implicitly when we write quantities like $I$ and $Q$ we are referring to $\langle I \rangle $ and $\langle Q \rangle$, where the brackets refer to time averages over intervals much longer than $2\pi / \omega$.

As a first example let us consider the simple case of a static magnetic field ${\bf B} = B_0 {\bf x}_2$ extending over a length $L$ in the ${\bf x}_3$ direction, such that $s \in (0,L)$.  We also take $\omega_{\rm pl}(s) = \omega_{\rm pl}$ to be independent of distance.  The point of this exercise is to gain familiarity with how competing effects contribute to $L_p$ before turning to the case of interest of conversion in stellar magnetospheres. Under the assumption that the photon wavelength is much smaller than the length $L$ ($2 \pi / \omega \ll L$), one may use a WKB approximation (see, {\it e.g.},~\cite{Raffelt:1987im}) to reduce the second-order axion-photon mixing equations to first-order mixing equations:
\es{eq:mixing}{
\left[ i \partial_s + \left(
\begin{array}{cc}
\Delta_{||} + \Delta_{\rm pl} & \Delta_B \\
\Delta_B & \Delta_a
\end{array}
\right)\right] \left(
\begin{array}{c}
A_{2} \\
a 
\end{array}
\right) = {\bf 0} \,,
}
with $A_2 = E_2 / (i \omega)$ the corresponding component of the vector potential in Weyl gauge ($A_0 = 0$), $\Delta_a =  - m_a^2 / \omega$, $\Delta_{\rm pl} = - \omega_{\rm pl}^2 / \omega$, $\Delta_B = g_{a\gamma\gamma} B_0 / 2$, and $\Delta_{||} = (7/2) \omega \xi$, with $\xi = (\alpha_{\rm EM} / 45 \pi) (B / B_{\rm crit})^2$, arising from the non-linear Euler-Heisenberg Lagrangian in strong-field quantum electrodynamics, with $B_{\rm crit} = m_e^2 /e \approx 4.41 \times 10^{13} \, \, {\rm G}$~\cite{2006physics...5038H}. 

Throughout this work we are interested in the weak mixing regime where the photon-to-axion conversion probabilities ($p_{\gamma \to a}$) and axion-to-photon probabilities ($p_{a\to \gamma}$) are much less than unity, so that we may work to leading non-trivial order in $g_{a\gamma\gamma}$.  We may then solve~\eqref{eq:mixing} in perturbation theory, treating the $\Delta_B$ mixing term as a perturbation, since without this term the mixing matrix in~\eqref{eq:mixing} is diagonal.  We consider the initial state, at $s = 0$, to be specified by the vector potential ${\bf A} = (A / \sqrt{2}) \left( a_1 {\bf \hat x}_1 + a_2 {\bf \hat x}_2 \right)$ for an arbitrary real $A$, where $a_1$ and $a_2$ are complex random variables that obey the relations: $\langle a_1 a_1* \rangle = \langle a_2 a_2* \rangle = 1$, with $\langle a_1 a_1 \rangle = \langle a_2 a_2 \rangle = \langle a_1 a_2 \rangle = \langle a_1 a_2* \rangle = 0$.  Referring to~\eqref{eq:stokes}, and recalling that all such quantities are subject to expectation values $\langle \dots \rangle$, we see that at $s = 0$ we have $I = A^2$, while $Q = U = V = 0$, implying that the initial state is unpolarized.  The perturbative solution to the equations of motion at $s > 0$ is then, up to unimportant phases and to second-order in perturbation theory,
\es{eq:pert}{
{\bf A}(s) = &{A \over \sqrt{2}} \Bigg[ a_1 {\bf \hat x}_1 + a_2 {\bf \hat x}_2 \Bigg( 1 - \\
& \int_0^s ds \Delta_B \int_0^{s'} ds'' \Delta_B e^{-i \int_0^{s''} ds''' \Delta_{\rm tr} }\Bigg)\Bigg] \,,
}
where in general~\eqref{eq:pert} would hold even if the mixing terms were $s$-dependent, though they are not in this simple example.  Note that we have defined $\Delta_{\rm tr} \equiv \Delta_{||} + \Delta_{\rm pl} - \Delta_a$.   Performing the integration in~\eqref{eq:pert} out to $s = L$ we find that
\es{eq:example_results}{
I &= A^2 \left( 1 - {\Delta_B^2 [1 -\cos(L \Delta_{\rm tr} ) ] \over \Delta_{\rm tr}^2 } \right) \,, \\
L_p &= {\Delta_B^2 \over \Delta_{\rm tr}^2} [1 - \cos(L \Delta_{\rm tr})] \,, \\
C_p &= 0 \,,
}
to leading non-trivial order in $\Delta_B$,
with the polarization angle $\chi = 0$.  Note that by the same logic the axion-to-photon conversion probability, for a pure initial axion state, is given by
\es{}{
p_{a \to \gamma} &= \left| \int_0^L ds' \Delta_B e^{-i \int_0^{s'} ds'' \Delta_{\rm tr} } \right|^2 \\
&= 2 {\Delta_B^2 \over \Delta_{\rm tr}^2} [1 - \cos(L \Delta_{\rm tr})] \,,
}
such that we may infer, at least for this example, that $L_p = p_{a\to\gamma} / 2$ to leading order in $\Delta_B$.  This should not be surprising in light of the physical picture of the underlying mechanism that produces the linear polarization.  The photons polarized in the ${\bf \hat x}_1$ direction are unaffected by the axion. However, those in the ${\bf \hat x}_2$ direction have a probability to convert to axions, $p_{\gamma \to a}$, which is equal to $p_{a \to \gamma}$. The photon survival probability is then $p_{\gamma \to \gamma} = 1 - p_{a\to\gamma}$.  Then, referring to~\eqref{eq:stokes} and~\eqref{eq:linear}, it is clear that $L_p = p_{a\to\gamma} / 2$. 

There are a few interesting points to be made about the expression for $L_p$.  If $| L \Delta_{\rm tr}| \ll 1$ then $L_p \approx {1 \over 2} \Delta_B^2 L^2$; the quadratic growth of $L_p$ with $L$ is related to the fact that the axion and photon remain in-phase during the mixing.  As $| L \Delta_{\rm tr}|$ becomes comparable to and greater than unity we begin to notice the different dispersion relations between the axion and photon over the distance $L$.  The difference of dispersion relations suppresses mixing.  Indeed, one surprising aspect of~\eqref{eq:example_results} is that if we assume $|\Delta_{||} | \gg |\Delta_a|, |\Delta_{\rm pl}|$ and $L | \Delta_{||}| \gg 1$, which would be the case appropriate for photons propagating over a large distance through a strongly magnetized region with low plasma density and an ultra-light axion in the spectrum, then the dependence of $L_p$ on $B_0$ is $L_p \propto 1/B_0^2$.  This is surprising because it suggests that when the Euler-Heisenberg term dominates $\Delta_{\rm tr}$, strong magnetic fields actually suppress mixing compared to weaker magnetic fields.

Let us now generalize the example above to consider dipole magnetic fields.  This is instructive because the magnetic fields surrounding many stars, such as the MWDs that are the main topic of this work but also the fields surrounding NSs and to a large extent main sequence stars as well, may be described -- at least to first approximation -- by dipole fields. Indeed, at distances far away from the star the field should approach that of a dipole, since the higher multipole field components fall off faster with distance. Let us assume that the star has a radius $R_{\rm star}$ such that unpolarized emission radiates from the surface and then propagates to infinity.  For the purpose of this example we will assume that ${\bf B}(s) = B_0 {\bf \hat x}_2 [R_{\rm star}  / (R_{\rm star} + s)]^3$, and we will compute $L_p$ with $s \to \infty$.  This magnetic field profile is that seen by radial emission at the magnetic equator, where ${\bf B}(s)$ remains perpendicular to the propagation direction for all $s$.  Moreover, we will make the assumption for this example that $|\Delta_{||}|$ dominates $\Delta_{\rm tr}$, which is the case appropriate for low-mass axions and low plasma densities.  In this case we may use~\eqref{eq:pert} to compute, to leading non-trivial order in $\Delta_B$,
\begin{widetext}
\es{eq:L_p_dipole_full}{
L_p \approx &1.4 \times 10^{-4} \left({g_{a\gamma\gamma} \over 10^{-12} \, \, {\rm GeV}^{-1} }\right)^2 \left( {B_0 \over 100 \, \, {\rm MG} } \right)^{2/5} \left( {1 \, \, {\rm eV} \over \omega } \right)^{4/5} \left( {R_{\rm star} \over 0.01 \, \, R_\odot } \right)^{6/5} \times \\
&{{\rm Abs}\left\{{\rm{Re}}\left[ (-1)^{2/5} e^{-i {7  \over 10} R_{\rm star} \xi_0 \omega} \left( \Gamma\left( {4 \over 5} \right) -\Gamma\left( {4 \over 5}, - {7 \over 10} i R_{\rm star} \xi_0 \omega \right) \right) \right] \right\} \over 0.022} \,,
}
\end{widetext}
with $\xi_0$ denoting the value at the surface such that
\es{}{
R_{\rm star} \xi_0 \omega \approx 9 \cdot 10^{-3} \left( {R_{\rm star} \over 0.01 \, R_{\odot} } \right) \left( {\omega \over 1 \, {\rm eV} } \right) \left( {B_0 \over 100 \, {\rm MG}} \right)^2 \,.
}
Note that when $R_{\rm star} \xi_0 \omega \ll 1$, which is a limit applicable to many MWD in this work, we may expand~\eqref{eq:L_p_dipole_full} to write
\es{eq:L_p_dipole_approx}{
L_p \approx &1.4 \times 10^{-4} \left({g_{a\gamma\gamma} \over 10^{-12} \, \, {\rm GeV}^{-1} }\right)^2 \left( {B_0 \over 100 \, \, {\rm MG} } \right)^{2} \\
&\times \left( {R_{\rm star} \over 0.01 \, \, R_\odot } \right)^{2} \,, \qquad R_{\rm star} \xi_0 \omega \ll 1 \,.
}
On the other hand, when $R_{\rm star} \xi_0 \omega \gg 1$, the  term appearing in the second line of~\eqref{eq:L_p_dipole_full} oscillates, with a typical magnitude around unity. That is, at very large magnetic field values, when the Euler-Heisenberg term dominates, $L_p \propto B_0^{2/5}$, while in the low-field limit the polarization scales more rapidly with magnetic field as $L_p \propto B_0^2$.  

There are a number of important points to be made regarding the formulae~\eqref{eq:L_p_dipole_full} and~\eqref{eq:L_p_dipole_approx}.  The MWDs in this work will have field values $\lesssim$1000 MG, and we will typically be considering energies $\omega \sim {\rm eV}$; thus, except in extreme cases -- such as high energies and high field values -- the Euler-Heisenberg term will not significantly affect $L_p$.  On the other hand, consider the searches in~\cite{Dessert_2019,Dessert:2021bkv} for hard $X$-rays arising from axion production in the cores of MWDs and converting to photons in the magnetospheres. In those works the typical axion energies are $\omega \sim {\rm keV}$, and thus we see that for the same MWDs the Euler-Heisenberg term is important to accurately describe the axion-to-photon conversion at those energies. On the other hand, consider an optical polarization signal arising from a strongly magnetic NS, with $R_{\rm star} \sim 10 \, \, {\rm km}$, $\omega \sim {\rm eV}$, and $B_0 \sim 10^{14} \, \, {\rm G}$.  Since $R_{\rm star} \xi_0 \omega \gg 1$ in that case we may infer that $L_p \approx 5 \times 10^{-5} (g_{a\gamma\gamma} / 10^{-12} \, \, {\rm GeV}^{-1})^2$.  Additionally, NS surface temperatures are typically much larger than an eV, with $\omega \sim 100 \, \, {\rm eV}$ being a more appropriate reference energy, which further suppresses $L_p$.  We thus arrive at the surprising conclusion that despite their lower magnetic field values, MWDs are more powerful probes of ultralight axions, with polarization probes, than NSs because the Euler-Heisenberg term suppresses axion-photon mixing in NS magnetospheres.

We may also use~\eqref{eq:L_p_dipole_approx} to verify that MWDs are more efficient at producing linear polarization than non-compact stars.  The Sun, for example, has a dipole magnetic field strength $B_0 \sim 10 \, \, {\rm G}$.  Thus, for unpolarized emission emanating from the non-active Sun we expect $L_p \sim 10^{-14} \left( g_{a\gamma\gamma} / 10^{-12} \, \, {\rm GeV}^{-1} \right)^2$.  Note that one of the most magnetized non-compact stars is HD 215441,
which hosts a dipole magnetic field of strength $\sim$30 kG and a radius $\sim$2 $R_\odot$~\cite{1960ApJ...132..521B}.  The axion-induced linear polarization fraction from this star would be $L_p \sim 5 \cdot 10^{-7} \left( g_{a\gamma\gamma} / 10^{-12} \, \, {\rm GeV}^{-1} \right)^2$, which is still subdominant compared to the MWD expectation.  

Indeed, we may make a general argument that, at least for $\omega \sim {\rm eV}$, strongly-magnetic MWDs are the optimal targets for axion-induced linear polarization searches.  
Stellar evolution approximately conserves magnetic flux across a surface far away from the star, such that the dipole field strength $B_f$ in a final stellar evolution stage is related to the initial field strength $B_i$ by $B_f \approx B_i (R_i / R_f)^2$, where $R_i$ ($R_f$) is the initial (final) stellar radius. Note that with this approximation we may re-scale the magnetic field of HD 215441 down to WD-radii stars ($R_{\rm star} \approx 0.01 \, R_\odot$) to estimate that the most strongly magnetized MWDs should have field strengths $B \sim 1000 \, \, {\rm MG}$, which is approximately correct.  Similarly, using this argument we may correctly infer that NSs can reach magnetic field values $\sim$$10^{15}$ G.
Using the flux conservation argument and assuming that we remain in the limit where we may neglect the Euler-Heisenberg term, we may relate the final-stage axion-induced polarization fraction $L_p^f$ to the initial-stage polarization fraction $L_p^i$: $L_p^f \approx L_p^i (R_i / R_f)^2$. This estimate suggests that more compact stars, such as MWDs, will be more efficient at producing axion-induced linear polarization than less compact stars.  On the other hand, this argument stops being true as soon as the Euler-Heisenberg term becomes important: at that point, the larger-radius star will produce a larger $L_p$.  As strongly-magnetic MWDs may achieve $R_{\rm star} \xi_0 \omega \sim 1$, we see that these are thus the optimal targets for axion-induced polarization studies.  For this reason, we will focus on these targets in this work.

So far we have neglected the possible effects of non-zero $\Delta_{\rm pl}$.  We now justify this approximation for MWD magnetospheres. 
The free electron density in the interstellar medium away from the Galactic Center may be as much as $n_e \sim 10^{-1}/{\rm cm}^3$, though in the outer parts of the Galaxy near the MWDs that are studied in this work it is typically lower~\cite{Cordes:2002wz}.  The plasma frequency associated with a free electron density $n_e = 10^{-1}/{\rm cm}^3$ is $\omega_{\rm pl}  = \sqrt{4 \pi \alpha_{\rm EM} n_e / m_e} \approx 10^{-11} \, \, {\rm eV}$, with $m_e$ the electron mass.  Referring back to {\it e.g.}~\eqref{eq:example_results}, the relevant dimensionless quantity to compute to assess the importance of the plasma mass term is $|R_{\rm star} \Delta_{\rm pl}| \approx 4 \times 10^{-9}$ for the above $n_e$ estimate, $\omega = 1$ eV, and $R_{\rm star} = 0.01 R_\odot$ appropriate for a WD.  Note that the plasma mass term would be important for $|R_{\rm star} \Delta_{\rm pl}| \gtrsim 1$.    Thus, even accounting for a significantly enhanced interstellar free-electron density near the MWD, it is unlikely that the $\Delta_{\rm pl}$ term would be important at optical frequencies.  On the other hand, within the MWD atmosphere the free-electron density may be significantly higher, perhaps as high as $n_e \approx 10^{17}/$cm$^3$~\cite{Gill_2011}.  
However, the MWD atmosphere is expected to have a density profile that falls exponentially with a characteristic scale height $\sim$100 m.  Considering that a typical WD radius is $\sim 7 \times 10^6$ m, we see that the atmosphere only extends non-trivially over a very small fraction of the stellar radius away from the surface.  The photon-to-axion conversion takes place continuously over a characteristic distance of order the MWD radius away from the stellar surface.  Thus, the effect of the atmosphere on the axion-induced contribution to $L_p$ is negligible.  More precisely, the effect of the atmosphere on the conversion probability is suppressed by the ratio of the MWD atmosphere thickness to the MWD radius; this ratio is $10^{-5}$.

In contrast to the axion-induced polarization signal, the standard astrophysical contributions to $L_p$ and $C_p$ arise solely within the atmosphere from anisotropic cyclotron absorption and bound-free transitions  \cite{lamb_sutherland_1974,angel1977magnetism}. 
In general, the degree of polarization is proportional to the optical depth of the atmosphere \cite{lamb_sutherland_1974}, so that the generation of astrophysical linear polarization is dominantly localized to within a characteristic scale height from the surface of the MWD.  We discuss the astrophysical contributions to the linear polarization in Sec.~\ref{sec:astro}, as they are a possible confounding background for the axion search.

Faraday rotation within the MWD magnetosphere and in the interstellar medium could in principle reduce the linear polarization fraction, though we estimate numerically that Faraday rotation is small (rotation angles up to $\sim$$10^{-10}$) for nearby MWDs with $B \lesssim 10^3$ MG and free electron densities of order those in the interstellar medium.

Returning to the axion-induced polarization signal, in the limit where we may neglect the Euler-Heisenberg term, we may also integrate~\eqref{eq:mixing} for a dipole magnetic field including the $\Delta_a$ term, but neglecting $\Delta_{\rm pl}$ for the reasons given above.  In this case, we find
\es{eq:L_p_large_ma}{
&L_p \approx 2 \times 10^{-8} \left({g_{a\gamma\gamma} \over 10^{-12} \, \, {\rm GeV}^{-1} }\right)^2 \left( {B_0 \over 100 \, \, {\rm MG} } \right)^{2} \\
&\times \left( {\omega \over 1 \, \, {\rm eV} } \right)^2 \left( {10^{-5} \, \, {\rm eV} \over m_a } \right)^4 \,, 
}
which is valid for $|r_0 \Delta_a| \gg 1$. Interestingly, $L_p$ is independent of $R_{\rm star}$ in the high axion mass limit. Nevertheless, the transition from the low mass to high mass region is dependent on $R_{\rm star}$, and in practice, the large-mass condition $|r_0 \Delta_a| \gg 1$ is satisfied for 
\es{}{
m_a \gg 1.7 \times 10^{-7}\, \, {\rm eV} \sqrt{\left({\omega \over 1 \, \, {\rm eV}} \right) \left( {0.01 \, \, R_\odot \over R_{\rm star} } \right) } \,.
}
Thus, we expect that MWD polarization studies to be insensitive to the axion mass for $m_a \lesssim 10^{-7}$ eV, while for masses much larger than this the sensitivity to $g_{a\gamma\gamma}$ should drop off quadratically with increasing $m_a$.
Next, we present the generalized mixing equations for non-radial trajectories including the Euler-Heisenberg Lagrangian. 

\subsection{General axion-photon mixing equations}
\label{sec:gen_mixing}

In this work we numerically solve the axion-photon mixing equations including the Euler-Heisenberg terms and also integrating over emission across the surface of the MWD.  That is, we assume that the MWD surfaces are isothermal (but see~\cite{2014Natur.515...88V}), such that the emission we see on Earth originates from across the full Earth-facing hemisphere of the MWD.  However, this means that photons that originate from across this surface that reach Earth will generically travel along non-radial trajectories, and this requires us to generalize the mixing equations in~\eqref{eq:mixing} to include mixing of the axion with both transverse modes:
\es{eq:mixing-gen}{
\left[ i \partial_s + \left(
\begin{array}{ccc}
\Delta_{11}  & \Delta_{12} & \Delta_{B_1} \\
\Delta_{12} & \Delta_{22} & \Delta_{B_2} \\
\Delta_{B_1} &  \Delta_{B_2} & \Delta_a 
\end{array}
\right)\right] \left(
\begin{array}{c}
A_{1} \\
A_{2} \\
a 
\end{array}
\right) = {\bf 0} \,.
}
Above, we assume that the photon travels along a straight trajectory in the direction ${\bf \hat s}$, with coordinate $s$, with ${\bf \hat x}_1$ and ${\bf \hat x}_2$ spanning the transverse directions.  We also neglect plasma terms because, as discussed above, they play a subdominant role.
The terms appearing in the mixing Hamiltonian in~\eqref{eq:mixing-gen} arise from axion-photon mixing, the Euler-Heisenberg Lagrangian, and the axion mass, and those that differ from the terms in~\eqref{eq:mixing} are defined by~\cite{Raffelt:1987im}
\es{}{
\Delta_{11} &= {2 \alpha_{\rm EM} \omega \over 45 \pi} \left[ {7 \over 4} \left(B_1 \over B_{\rm crit} \right)^2 + \left(B_2 \over B_{\rm crit} \right)^2  \right] \, \\
\Delta_{22} &= {2 \alpha_{\rm EM} \omega \over 45 \pi} \left[ {7 \over 4} \left(B_2 \over B_{\rm crit} \right)^2 + \left(B_1 \over B_{\rm crit} \right)^2  \right] \, \\
\Delta_{12} &= {3 \over 4} {2 \alpha_{\rm EM} \omega \over 45 \pi} \left(B_1 B_2 \over B_{\rm crit}^2 \right) \,, \quad \Delta_{B_i} = {1 \over 2} g_{a\gamma\gamma}B_i  \,, 
}
with \mbox{$i = 1,2$} in the last line.  Above, $B_1$ and $B_2$ are the magnetic field values in the transverse directions, and they are generically functions of $s$.

When applying the formalism above to predict the axion-induced $L_p$ from a MWD, we begin by discretizing the surface of the hemisphere of the Earth-facing MWD.  We consider initially unpolarized emission from each surface element propagating in the ${\bf \hat x}_3$ direction, with the final $A_1$ and $A_2$ being the appropriate sum of the contributions from the different surface elements.  This is accomplished by letting the initial vector potential of each surface element $i$ be labeled as ${\bf A}^i = (A^i / \sqrt{2}) \big( a_1^i {\bf \hat x}_1 + a_2^i {\bf \hat x}_2 \big)$, where the $a_1^i$ and $a_2^i$ are uncorrelated random variables such that $\langle a_1^i {a_1^j}^* \rangle = \delta^{ij}$ with all other correlators vanishing.  We adjust the normalization parameter $A^i$ such that $A^i \propto \sqrt{0.7 + 0.3 \cos \theta_i}$, with $\theta_i$ being the angle between the normal vector to the sphere at pixel $i$ and the ${\bf \hat x}_3$ axis.  This scaling reproduces the limb darkening law for the intensity adopted in~\cite{Euchner:2002qv}, who confirmed this scaling through radiative transfer calculations. 

\subsubsection{Magnetic white dwarf magnetic field models}
'
The magnetic field profile around a compact star will generically approach that of a dipole configuration far away from the stellar surface, since higher-harmonic contributions to the vacuum solutions to the Maxwell equations fall off faster with radius.  In this work, we will consider both pure dipole profiles and profiles containing higher harmonic modes, which have been fit to luminosity and circular polarization data from specific MWDs.  The dipole solution may be written as 
\es{eq:dipole}{
{\bf B}({\bf r}) = {B_p \over 2} \left( {R_{\rm star} \over r} \right)^3 \left[ 3 {\bf \hat r} ( {\bf \hat m} \cdot {\bf \hat r} )- {\bf \hat m} \right] \,,
} 
where ${\bf \hat m}$ points along the polarization axis in the direction of the magnetic north pole and ${\bf \hat r}$ is the position unit vector, with distance $r$ from the center of the star.  The field strength $B_p$ is the polar value at the surface of the star.  

The general solution to the Maxwell equations in vacuum may be written in terms of spherical harmonics; the associated magnetic scalar potential $\psi$, defined such that ${\bf B} = -{\bf \nabla} \psi $, is given by 
\es{eq:harm}{
\psi = - R_{\rm star}\sum_{\ell = 1}^{\infty} \sum_{m = 0}^\ell &\left( {R_{\rm star} \over r} \right)^{\ell + 1} \left[ g_{\ell}^m \cos m \phi \right. \\
&\left. + h_{\ell}^m \sin m \phi \right] P_\ell^m (\cos \theta) \,,
}
where the coefficient $g_\ell^m$ and $h_\ell^m$ have dimensions of magnetic field strength.  The angle $\theta$ is the angle away from the polarization axis ${\bf \hat m}$, such that ${\bf \hat m} \cdot {\bf \hat r} = \cos \theta$, and the angle $\phi$ is the rotation angle about ${\bf \hat m}$.  The $P_{\ell}^m$ are the associated Legendre polynomials.  Note that the terms in~\eqref{eq:harm} at $\ell = 1$ are simply those in~\eqref{eq:dipole} for the dipole configuration.  Ref.~\cite{2003ASIB..105..175J} provides a fit of the harmonic solution in~\eqref{eq:harm} to MWD circular polarization and spectra data for Grw+70$^\circ$8247 up through $\ell \leq 4$; we will make use of this fit later in this work.

It is convenient to define an inclination angle $i$ that is the angle between the magnetic axis ${\bf \hat m}$ and the direction towards Earth.  For definiteness, throughout this work we define the coordinate system centered at the MWD center with ${\bf \hat z}$ pointing towards the Earth and with ${\bf \hat m} = \cos i {\bf \hat z} + \sin i {\bf \hat y}$.  Note that for a dipole field configuration the linear polarization must vanish as $i \to 0$, since in this limit there is no preferred direction for the linear polarization to point. 

\subsection{Astrophysical contributions to the linear polarization}
\label{sec:astro}

Astrophysical mechanisms exist within the MWD atmospheres for polarizing the outgoing radiation.  Like the axion mechanism that is the focus of this work, the astrophysical mechanisms also rely on the polarizing effects of the magnetic field.  Here, we overview the calculation of the astrophysical polarization, as astrophysical emission serves as a background contribution in the axion searches that we discuss later in this work.    
As we will see one crucial difference between the two sources of linear polarization is that the astrophysical mechanisms lead to strong wavelength dependence of the polarization fraction, while the axion-induced polarization depends less strongly on wavelength.  This difference helps constrain the axion-induced linear polarization fraction even in the presence of an unconstrained astrophysical polarization fraction, which in principle could partially interfere with the axion signal at certain wavelengths.

In what follows we assume that the MWD atmosphere is composed primarily of hydrogen, which is the case for the MWDs we consider in this work. The bound electrons in the MWD atmosphere can be considered in the Paschen-Back regime, where the Hamiltonian is given by
\es{eq:HPB}{H = \dfrac{p^2}{2m_e} - \dfrac{\alpha_{\rm EM}}{r}  + \dfrac{1}{2}\Omega_C L_z + \dfrac{1}{8}m_e\Omega_C^2 r^2  \sin^2{\theta} \,,} 
with the third term accounting for the linear Zeeman effect and the fourth term the quadratic Zeeman effect.  The electron mass is $m_e$, the cyclotron frequency is $\Omega_c = e B / m_e$, $r$ is the atomic radial distance, and $\theta = 0$ points along the magnetic field.  At the fields under consideration $B\gtrsim 100$ MG, the quadratic Zeeman effect is important or dominant. However, in this work we use an approximation for fields $B \lesssim 100$ MG to model the astrophysical linear polarization, given by Ref.~\cite{lamb_sutherland_1974} and Ref.~\cite{1992A&A...265..570J}. The reason is that the bound-free transition cross sections have not yet been computed with sufficient resolution for the modeling of MWD polarization at high field values. Recent advances in solving the Hamiltonian of~\eqref{eq:HPB} have led to numerical cross sections for a limited number of these transitions, but they were not reported on a fine enough grid of magnetic fields strengths for astrophysical modeling~\cite{1995A&A...298..193M,2007ApJ...667.1119Z,2021ApJS..254...21Z}.

Here, we first describe the generation of polarization for low fields, where the quadratic Zeeman effect is negligible. There are two main astrophysical processes that contribute to continuum linear and circular polarization of MWD starlight: (1) the ionization of a bound electron in a hydrogen atom (bound-free polarization) and (2) the absorption of a photon by an ionized electron (free-free polarization) \cite{lamb_sutherland_1974}. Bound-bound transitions of the hydrogen atom can produce localized features in the MWD spectra, and the observation of these features are used to estimate the surface magnetic fields of MWD, as the bound state energies of the hydrogen atom have been solved. Bound-bound transitions can also contribute to the polarization continuum, but these effects are washed-out by the large variation in the field on the MWD surface. We discuss the bound-bound transitions further in the context of SDSS J135141 in Sec.~\ref{sec:J1351Abs}.

The MWD starlight is produced unpolarized deep within the atmosphere as blackbody radiation. The polarization is generated as the light propagates through the thin atmosphere and ionizes bound electrons and scatters on free electrons. Because the atmosphere is thin compared to the coherence length of the magnetosphere, to a good approximation the magnetic field is constant throughout the atmosphere at a given point on the surface of the MWD. This surface magnetic field preferentially selects a direction for the absorption to occur, which polarizes the blackbody radiation. The bound-free transitions must satisfy the dipole selection rules $q = 0, \pm 1$, where $q$ is the difference between the initial and final magnetic quantum numbers, $m_i$ and $m_f$, respectively, of the transition. The transitions with $q = \pm 1$ preferentially absorb photons polarized perpendicular to the magnetic field and therefore polarizes the starlight parallel to the magnetic field. On the other hand, the transitions with $q = 0$ preferentially absorb photons of the opposite polarization, so that these transitions polarize the starlight perpendicular to the magnetic field. To determine the overall effect of bound-free absorption, there is a competition between these two terms. Over the majority of the photon energy range, the $q = \pm 1$ transitions are stronger such that the starlight is polarized parallel to the magnetic field. Only for photon energies near the hydrogen absorption edges does the polarization flip so that the linear polarization points perpendicular to the magnetic field. Finally, for free-free absorption, light is preferentially absorbed in the plane perpendicular to the magnetic field because the cyclotron motion of the free electrons restricts them to this plane, and therefore this absorption polarizes the light parallel to the magnetic field. If the axion-induced polarization is perpendicular to the astrophysical polarization direction then the two signals may partially destructively interfere.

Quantitatively, the effect of the bound-free and free-free absorption may be captured though the transfer equation describing the evolution of the photon polarization state matrix  (effectively a photon density matrix),
\begin{align}
\mathcal{F} = 
\begin{pmatrix}
    E_1 
    \\
    E_2
\end{pmatrix}
\begin{pmatrix}
    E_1^* & E_2^*
\end{pmatrix} = \frac{1}{2} S^\mu \sigma_\mu \, ,
\end{align}
where $S^\mu = (I, Q, U, V)$ and $\sigma_\mu = (\mathds{1}, \sigma_z, \sigma_x, \sigma_y)$ are the Stokes and Pauli vectors, respectively. 
In the anisotropic atmospheric plasma of the MWD, the transfer equations take the form \cite{lamb_sutherland_1974},
\begin{align}
\label{eq:transferEq}
\frac{d \mathcal{F}}{ds} = - \frac{1}{2} (T \mathcal{F} + \mathcal{F} T^{\dagger}) + \mathcal{E} \, ,
\end{align}
where the transfer matrix $T = \mathcal{K} - 2 i \mathcal{R}$ describes  absorption ($\mathcal{K})$ and refraction ($\mathcal{R}$), while $\mathcal{E}$ describes  emission. Equation~\eqref{eq:transferEq} can be solved analytically under the approximation that the initially unpolarized blackbody radiation emanating from the MWD experiences a constant magnetic field while traversing the thin, cold, atmosphere.  As shown in \cite{lamb_sutherland_1974}, under these assumptions, the solution to \eqref{eq:transferEq} as expressed in terms of the final polarization state of starlight leaving the MWD atmosphere of thickness $\delta s$ is given in terms of the Stokes parameters by \cite{lamb_sutherland_1974}
\es{eq:stokesAstrophysical}{
\begin{aligned}
I &= 1-\frac{\delta s}{2}  {\rm tr}(\mathcal{K}) \,, & Q &= -\frac{\delta s}{2}  {\rm tr}(\sigma_{z} \mathcal{K}) \, , \\
U &= -\frac{\delta s}{2}  {\rm tr}(\sigma_{x} \mathcal{K}) \,, &  V &= -\frac{\delta s}{2}  {\rm tr}(\sigma_{y} \mathcal{K}) \,.
\end{aligned}
}

For dipole transitions like bound-free and cyclotron absorption, $\mathcal{K}$ is diagonal in the complex spherical basis with matrix elements
\begin{align}
    \label{eq:generalK}
    \mathcal{K}_q(\omega) = n \sigma_q(\omega) \, ,
\end{align}
where $n$ is the number density of the absorbing species and $\sigma_q$ the associated frequency-dependent cross-section, with $\omega$ the radiation frequency.

The astrophysical linear polarization follows from~\eqref{eq:stokesAstrophysical} and~\eqref{eq:generalK} and is given by 
\begin{align}
    \label{eq:astrophysicalPolarization}
    L_{p, \rm astro} = \frac{|Q|}{I} = \frac{\delta s}{4}|2 \mathcal{K}_0 - \mathcal{K}_+ - \mathcal{K}_-| \sin^2 \theta  \, ,
\end{align}
since $U=0$ in this basis. As in~\eqref{eq:HPB}, $\theta$ is the angle between the surface magnetic field and the light propagation direction, and $\mathcal{K}$ in general includes bound-bound, bound-free, and free-free absorption contributions, although we do not consider bound-bound transitions. 

Note that~\eqref{eq:astrophysicalPolarization} holds for any MWD magnetic field strength. 
However, for MWDs with high fields where the linear Zeeman effect breaks down ($B \gtrsim 100$ MG), the bound-free absorption cross-section become difficult to calculate. In this work we use an approximation that is common in the literature.
For bound-free collisions where the quadratic Zeeman effect is unimportant ($B \lesssim 100$ MG), $\mathcal{K}_q$ can be calculated analytically under the approximation that the wavefunction of the bound electron is unaffected by the perturbing external magnetic field while its energy shifts linearly by $m_i\Omega_C$. Under these approximations, the bound-free absorption cross-section was derived first in~\cite{lamb_sutherland_1974}.

We use the improved approximation~\cite{1992A&A...265..570J} that accounts for the energies of the hydrogen absorption edges $\epsilon_{nlmq}$ as a function of magnetic field, $\epsilon_{nlmq} \equiv E_{nlm}(B) + \Theta\left(m_f\Omega_C\right)$ for $\Theta$ the Heaviside step function. The first term accounts for the fact that the bound state energies of hydrogen in the quadratic Zeeman regime depend on all three quantum numbers \{$n$,$l$,$m$\} and the magnetic field strength $B$, because the Hamiltonian of~\eqref{eq:HPB} breaks spherical symmetry. These bound state energies $E_{nlm}(B)$ are tabulated in~\cite{2014ApJS..212...26S}. We also account for the quantization of the free electrons into Landau levels, which yields the second term. Then the bound-free absorption coefficients are given by
\es{eq:absorpMatrixBF_Jordan}{
\mathcal{K}_{q,\rm bf}(\omega) = n_{\rm H}{\omega \over \omega - q \Omega_C}\sum_{nlm}^{n\leq 4} & \exp\left({-E_{nlm}(B) \over T}\right) \times \\
& \begin{cases}
\sigma_n^{\rm bf}(\omega - q \Omega_C), & \omega \geq \epsilon_{nlmq}  \\
0, & \omega < \epsilon_{nlmq}
\end{cases}  \,.
}
We weight the states with the Boltzmann factor, under the assumption of the fixed surface temperature $T = 15000$ K, appropriate for the MWDs we consider in this work. $\sigma_n^{\rm bf}(\omega) \propto n^{-5}\omega^{-3}$ is the cross section for a photon of energy $\omega$ to ionize an electron of principal quantum number $n$ at zero magnetic field. The dependence on $\omega - q \Omega_C$ is derived in the linear Zeeman regime. For the optical spectra we consider in this work, we only need to consider $n \leq 4$. 

The free-free absorption matrix is proportional to the cyclotron absorption cross-section
\es{eq:absorpMatrixFF}{
\mathcal{K}_{q,\rm ff}(\omega) = 
\begin{cases}
n_e \sigma^{\rm ff} & q = +1\\
0 & q \neq +1
\end{cases}
,}
where $n_e$ the number density of free electrons. We take the cyclotron absorption cross-section $\sigma^{\rm ff}$ as given in \cite{lamb_sutherland_1974}.  Only the $q = 1$ component is nonzero due to selection rules that enforce energy and angular momentum conservation along $\mathbf{B}$ \cite{lamb1972line}, and this cross section is strongly peaked around $\omega = \Omega_C$.

At low magnetic fields $B\lesssim100$ MG, the cyclotron frequency is much smaller than the optical frequencies, so that we do not need to consider cyclotron absorption contributions to the atmospheric opacity. Thus, only the bound-free absorption cross-section \eqref{eq:absorpMatrixBF_Jordan} contributes to the polarization. Furthermore, the hydrogen absorption edges are close to their zero-field values $13.6$ eV/$n^2$. Then, for energies far away from the absorption edges \eqref{eq:astrophysicalPolarization} reduces to~\cite{lamb_sutherland_1974} 
\es{eq:astro-LP-low-field}{
L_{p,{\rm astro}}(\omega) \propto  \dfrac{\Omega_C^2}{\omega^5} \sin^2{\theta} \, .} 
The proportionality constant of \eqref{eq:astro-LP-low-field} depends on the line-of-sight integrated bound electron density in the MWD atmosphere. Since $\Omega_C \propto B$, we see that in this regime the astrophysical linear polarization scales as the transverse magnetic field strength squared like that induced by the axion. 
However, the astrophysical polarization points parallel to the magnetic field while the axion-induced polarization points perpendicular to the field, which means that the two contributions may partially cancel each other depending on their relative magnitudes.

By contrast, even at low magnetic fields, the linear polarization displays strong localized features near the absorption edges. The linear polarization becomes much larger in magnitude and switches direction blueward of the edge so that it points perpendicular to the magnetic field, in the same direction as the axion-induced polarization.

However, in this work we consider MWDs with large magnetic fields $B\gtrsim100$ MG. In this case, the cyclotron frequency enters the optical, so that we must include the cyclotron absorption contribution to the linear polarization. The bound-free absorption also becomes more complex than at lower fields. The absorption edges cover nearly the entire optical spectrum. Furthermore, the hydrogen bound state energies depend strongly on the magnetic field strength, and the magnetic field strength on the surface of the MWD may span more than a factor of two, which additionally broadens the absorption edge features. Under the approximation used in this work~\eqref{eq:absorpMatrixBF_Jordan}, which assumes the bound-free cross section is simply that at zero-field shifted by $q\Omega_C$, we find that most of the linear polarization spectrum is dominated by the absorption edge features rather than by the simple power law scaling of~\eqref{eq:astro-LP-low-field}. The exact cross sections have been previously computed numerically for a limited number of transitions~\cite{1995A&A...298..193M,2007ApJ...667.1119Z,2021ApJS..254...21Z}. In these results there are additional oscillatory features near Landau thresholds, where the photon energy matches the energy difference between a Rydberg bound state and a Landau level. We thus expect that the eventual incorporation of the numerical cross sections into MWD linear polarization calculations will introduce additional features in the spectra due to these resonances, although these features will be smeared out due to the range of field strengths on the MWD surface.

At still higher magnetic fields $B \gtrsim 5000$ MG, the situation becomes less complicated. The quadratic Zeeman term dominates the Coulomb term in~\eqref{eq:HPB}. The approximation that the Coulomb field is a perturbation on the background magnetic field becomes more appropriate, and in this limit, we find, following \cite{landau2013quantum}, that $\sigma^{\rm bf}$ scales as $\omega^{-3}$ away from absorption edges as in the low-field case. 

Despite the uncertainties described above, essentially any energy dependence in the astrophysical polarization is sufficient to distinguish it from the axion-induced polarization for the purpose of setting an upper limit on the axion-induced polarization contribution, which is approximately energy independent, given spectropolarimetric data. 
As discussed further in Sec. \ref{sec:J135_data}, this is because given some amount of energy dependence in the astrophysical background, the axion and astrophysical contributions would not completely destructively interfere across the full analysis energy range.
On the other hand, in order to claim evidence for an axion signal, the astrophysical linear polarization signal should be better understood in the high-field regime.  This is because without a full understanding of how the astrophysical polarization emerges in the high field regime, one cannot be confident that a putative signal arises from axions and not the imprecisely known astrophysical polarization mechanisms.

\section{Upper Limits on $g_{a\gamma\gamma}$ from Magnetic White Dwarfs}

In this section we apply the formalism developed in the previous section to set upper limits on $|g_{a\gamma\gamma}|$ from linear polarization data towards the MWDs  SDSS J135141 (Sec.~\ref{sec:J135141}) and GRW$+$70$^\circ$8247 (Sec.~\ref{sec:GRW}).  These MWDs are unique in that they have strong but well-characterized magnetic field profiles in addition to dedicated linear polarization data.  We discuss additional MWDs that are promising but have somewhat incomplete data at present in Sec.~\ref{sec:other}.

\subsection{SDSS J135141}
\label{sec:J135141}

The MWD SDSS J135141 has one of the largest magnetic fields of all known MWDs.  Ref.~\cite{2009A&A...506.1341K} measured the polar magnetic field strength in the context of the dipole model to be $B_p =761.0 \pm 56.4$ MG, with an inclination angle $i  = 74.2^\circ \pm 21.7^\circ$.\footnote{Note that Ref.~\cite{2009A&A...506.1341K} also considered an offset dipole model, but we do not consider this model here.}
In the below analysis we consider 
the dipole model, 
and we compute the 95\% upper limit on $|g_{a\gamma\gamma}|$ considering the range of allowable magnetic field parameters.  In particular, we take our fiducial limit to be the weakest one across the range of allowable magnetic field parameters, allowing the parameters to vary within their 1$\sigma$ ranges, while we calculate the 95\% confidence level statistical upper limit on the data itself.  

\subsubsection{Absorption lines and magnetic field model}
\label{sec:J1351Abs}

In this section we overview the determination of the SDSS J135141 magnetic field strength. To date, this determination has been made only through spectra rather than polarimetry, although the addition of polarimetery would be beneficial to further constraining the magnetic field profile on the surface. The spectrum of a MWD is that of a thermal distribution at the temperature of the MWD surface, but with absorption features at wavelengths at which bound-bound transitions occur in the atmosphere. The transition wavelengths are very strongly dependent on the local magnetic field; therefore, the absorption lines are broadened by the range of magnetic field strengths on the MWD surface. In many cases the features are entirely washed out because the transition wavelengths are highly dependent on the local magnetic field, but a few transitions are nearly stationary because they encounter local extrema. The primary method for determining the magnetic field strength of MWDs is to search for these stationary features in the spectrum. The bound-bound transitions and dipole transition strengths of the hydrogen atom in a strong magnetic field are given in Ref.~\cite{1994asmf.book.....R}.

\begin{figure}[!htb]
\begin{center}
\includegraphics[width=0.49\textwidth]{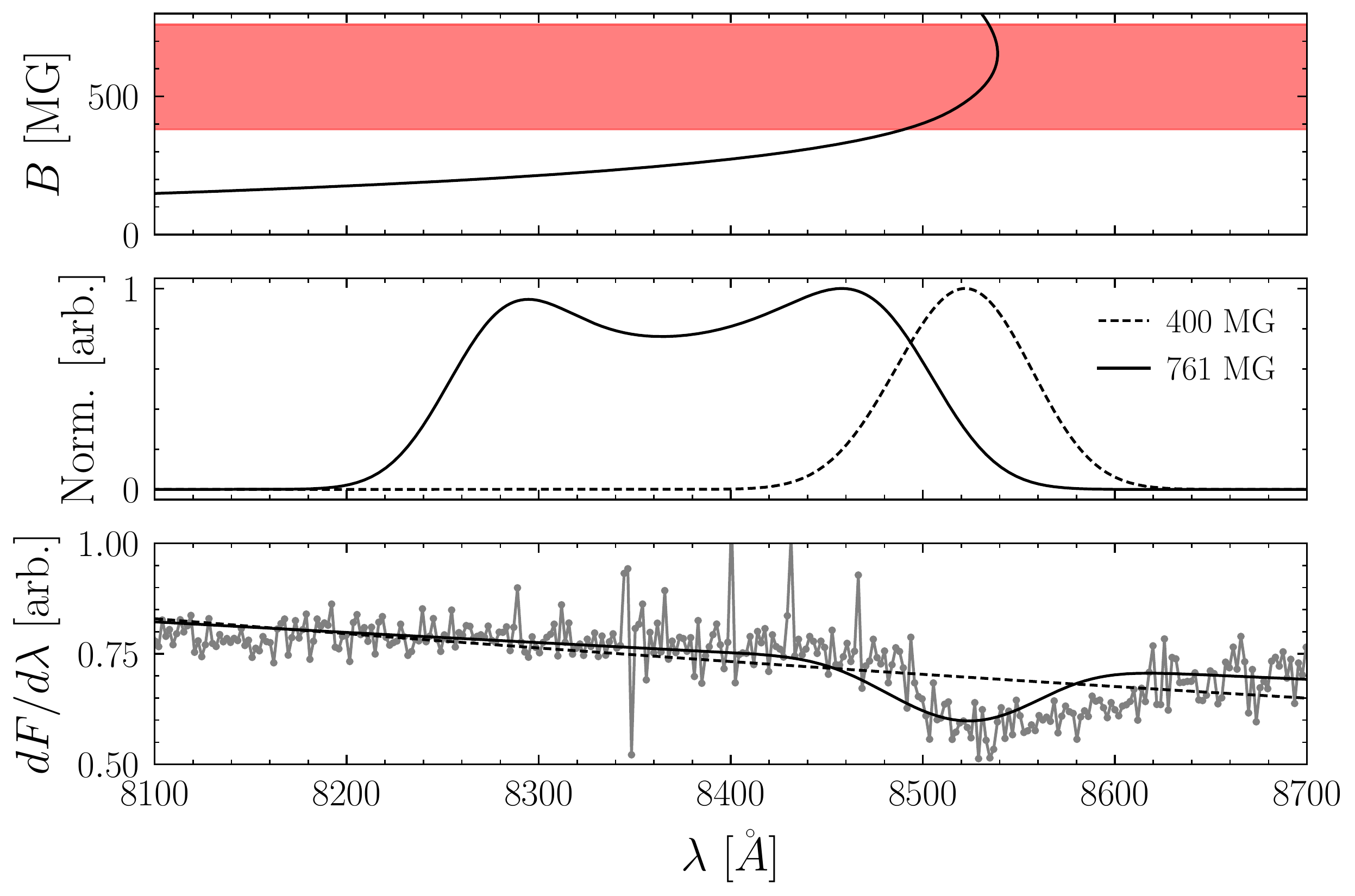}
\caption{
\label{fig:HighField} (Top) The wavelength of the $3d_{-1}-2p_0$ absorption line as a function of magnetic field. The red shaded region indicates the range of field strengths present on the surface, assuming the best-fit dipole field of 761 MG from~\cite{2009A&A...506.1341K}. (Middle) In solid black is the $3d_{-1}-2p_0$ line template for a 761 MG dipolar field; in dashed black for 400 MG. (Bottom) The flux of SDSS J135141 as measured by SDSS DR7 (gray). In solid black is the best fit spectrum assuming a 761 MG dipole field. In dashed black is the best fit spectrum assuming a 400 MG dipole field.
}
\end{center}
\end{figure}

In Fig.~\ref{fig:HighField} we show the wavelength dependence as a function of magnetic field for the stationary bound-bound $3d_{-1}-2p_0$ transition in the upper panel. The transition is nearly stationary around across the full range of field strengths present on the surface of SDSS J135141, assuming the 761 MG dipolar field. In the middle panel, we show the expected line templates for two cases (i) the best-fit dipolar field of 761 MG~\cite{2009A&A...506.1341K} and inclination angle $i = 74.2^\circ$, and (ii) a dipolar field of 400 MG with best-fitting $i$ for that field strength. To compute these templates, we histogram the wavelengths of the transition on the visible hemisphere of the MWD and weight each contribution by the dipole transition strength.  We also incorporate the limb darkening law mentioned previously from Ref.~\cite{Euchner:2002qv}, which weights the intensities between pixels on the sphere such that $I \propto 0.7 + 0.3 \cos \theta$, with $\theta$ the angle of the normal to the ${\bf \hat x}_3$ axis that points towards Earth. Note that due to the symmetry present in a dipole field, it is only the limb darkening rule that changes the spectral shape of the template with inclination angle $i$. The template is then convoluted with a Gaussian that has standard deviation $\sigma_{\rm stark}$.  This broadening is due to the Stark effect, accounting for the electric field that is also present on the MWD surface, and is the dominant broadening effect for these lines. 
We treat $\sigma_{\rm stark}$ as a nuisance parameter that is determined by maximum likelihood estimation.

For the 761 MG case, the absorption line appears at approximately the same location across the entire hemisphere, so that the resulting feature is highly localized around 8530 {\AA}. On the other hand, if the MWD had a lower field strength of 400 MG, the feature would be significantly broadened because the transition is not stationary at those field strengths, and additionally the feature would appear at shorter wavelengths $\sim 8200-8600$ {\AA}. In the lower panel, we fit expected flux models for each case to the SDSS data~\cite{2009A&A...506.1341K}. The models are a power law background with free index and normalization with the multiplicative absorption template as shown in the middle panel.
For the 761 MG case, we see that the model prefers an absorption line, indicating that the 761 MG dipole is a reasonable fit to the data. On the other hand, for the 400 MG case, the fit finds no evidence for a line.  Following a similar procedure SDSS J135141 was determined to have a $761.0 \pm 56.4$ MG field~\cite{2009A&A...506.1341K}, although that work fit to the broad-band flux spectra over a much larger wavelength range encompassing many absorption lines.  In fact, Ref.~\cite{2009A&A...506.1341K} did not include the wavelength range shown in Fig.~\ref{fig:HighField} in their fit; the fact that their best-fit model from lower wavelengths also explains the $3d_{-1}-2s_0$ absorption line feature provides non-trivial evidence that the magnetic fields on the surface of the MWD are $\sim$400-700 MG.

\subsubsection{Polarization data}
\label{sec:J135_data}
\begin{figure}[!t]
\begin{center}
\includegraphics[width=0.48\textwidth]{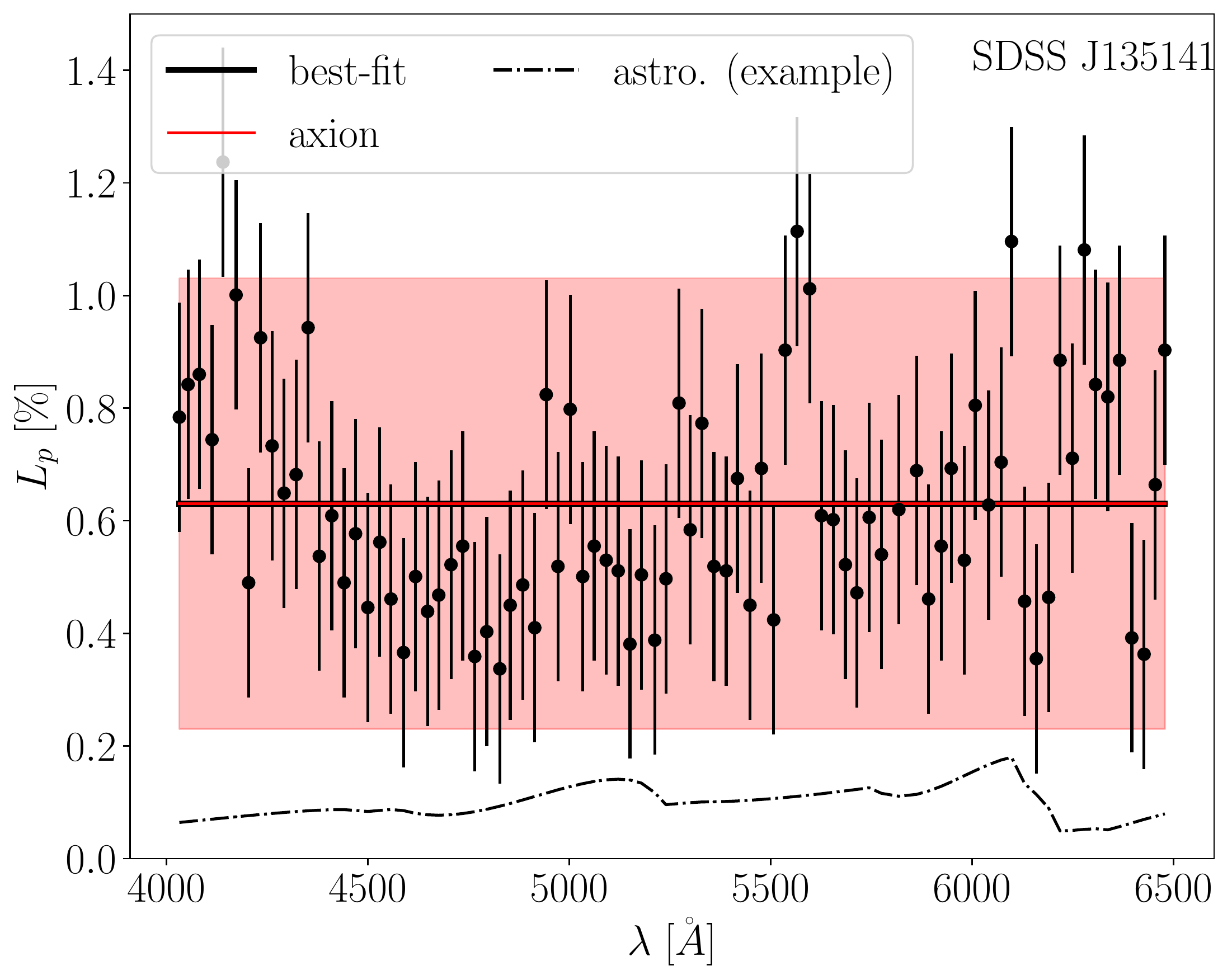} 
\caption{
\label{fig:J135_data}
The linear polarization data as a function of wavelength towards the MWD SDSS J135141 as observed by~\cite{2019ASPC..518...93P} with the SAO 6-m telescope.  We use a Gaussian likelihood to fit a model to the data with three components: (i) the axion signal, (ii) the astrophysical background, and (iii) an instrumental systematic contribution.  We assume that the axion signal and the instrumental systematic are wavelength-independent, while the astrophysical background depends on wavelength as described in Sec.~\ref{sec:J135_data}.  The axion signal and the instrumental systematic contributions would be completely degenerate, given that the systematic normalization  parameter can take either sign, but for the prior on the systematic nuisance parameter.  The best fit model, along with the axion contribution to that model, are illustrated, along with the best-fit statistical uncertainties on the data; the statistical uncertainty is treated as a hyperparameter that is determined by maximum likelihood estimation.  The red band illustrates the allowed axion contribution at 1$\sigma$ confidence.  At the best-fit point the astrophysical normalization is zero.  Still, we illustrate  the astrophysical  linear polarization model, with an arbitrary normalization. 
}
\end{center}
\end{figure}

The linear polarization of SDSS J135141 was measured in 2007 by~\cite{2019ASPC..518...93P} using the Special Astrophysical Observatory (SAO) 6-m telescope with the Spectral Camera with Optical Reducer for Photometric and Interferometrical Observations (SCORPIO) focal reducer~\cite{Afanasiev:2005yg}.  Across the wavelength range 4000~{\AA} to 6500~{\AA} the linear polarization fraction was measured to be $L_p = 0.62\% \pm 0.4\%$.  The uncertainty on $L_p$ is dominated by the systematic uncertainty, arising from effects such as scattered light and ghosts~\cite{Afanasiev:2005yg}, though the exact systematic uncertainty accounting that goes into the $L_p$ measurement is not detailed in~\cite{2019ASPC..518...93P}.  The linear polarization fraction data from~\cite{2019ASPC..518...93P} is reproduced in Fig.~\ref{fig:J135_data}.

An upper limit on the average axion-induced polarization fraction over the wavelength range $L_{p,{\rm axion}}$ may be estimated by the requirement that  axions not overproduce the observed polarization, which at 95\% confidence and assuming Wilks' theorem implies $L_{p,{\rm axion}} \lesssim 0.62 \% + \sqrt{2.71} \times 0.4 \% \approx 1.28\%$~\cite{Cowan:2010js}.  This upper limit is very close to that we will derive below making use of the wavelength dependent data and incorporating the astrophysical background model.  This point illustrates that 
the astrophysical polarization contribution is not a limiting background for constraining the axion-induced polarization, at least for this example.  This is fundamentally because the astrophysical background and the axion signal are polarized in the same direction over the wavelength range relevant for this search.  Our polarization upper limit is also consistent with that found in~\cite{1994ApJ...421..733L}, who performed spectropolarimetric observations of the MWD using the Steward Observatory 2.3 m telescope in 1993 and state that the linear polarization of SDSS J135141 in the wavelength range 4100 {\AA} to 7280 {\AA} was found to be less than 1\%, though the confidence level of that statement is not given in~\cite{1994ApJ...421..733L}.

To analyze the wavelength dependent data, we adopt a Gaussian likelihood function that incorporates the systematic uncertainty in a straightforward way, though the following analysis could likely be improved in the future with a better understanding of the origin of the systematic uncertainty.  The likelihood we adopt is given by
\es{eq:LL}{
p({\bf d} | {\mathcal M}, {\bm \theta}) = \left(\prod_i {1 \over \sigma} e^{-(d_i - L_p({\bm \theta}))^2 \over 2 \sigma^2} \right) e^{-A_{\rm sys}^2 \over 2 \sigma_{\rm sys.}^2} \,,
}
where we leave off unimportant numerical normalization factors and where $i$ labels the wavelength bins (there are $83$ different wavelength bins, as illustrated in Fig.~\ref{fig:J135_data}).  The data ${\bm d}$, with entries $d_i$, are the observed polarization values, while the model ${\mathcal M}$ has parameters ${\bm \theta} = \{A_{\rm axion},A_{\rm astro},A_{\rm sys},\sigma\}$.  The signal parameter $A_{\rm axion}$ controls the normalization of the axion-induced polarization and, physically, is a proxy for $g_{a\gamma\gamma}$, at fixed $m_a$.  The parameter $A_{\rm astro}$ controls the amplitude of the unknown astrophysical background.  The instrumental ({\it e.g.}, systematic) contribution to the polarization is characterized by the nuisance parameter $A_{\rm sys}$.  The parameter $\sigma$ may be interpreted as the uncorrelated statistical uncertainty on the linear polarization data.  We treat $\sigma$ as a hyperparameter that is determined by maximum likelihood estimation.  

Both the astrophysical and axion contributions to the polarization in principle have non-trivial wavelength dependence; in the axion case, the wavelength dependence is found by numerically solving the axion-photon mixing equations, while for the astrophysical contribution we use~\eqref{eq:astrophysicalPolarization}. For all of the magnetic field models, only bound-free absorption contributes, as the cyclotron wavelength is not in the wavelength range of the data. We compute the Stokes parameters by averaging them over $\sim10^5$ points on the MWD surface in each wavelength bin. The full list of absorption edges and associated wavelength ranges that contribute to features in the astrophysical linear polarization model are given in Tab.~\ref{tab:AbsEdges}. Accounting for the uncertainty on the magnetic field strength and orientation, the edges may shift by $\sim 100$ {\AA}.
\begin{table}[]
\centering
\renewcommand{\arraystretch}{1.3}
\begin{tabularx}{0.34\textwidth}{YYY}
\hline
$nl_{\rm m}$ & $q$      & $\epsilon_{nlmq}$ ({\AA}) \\ \hline
\renewcommand{\arraystretch}{1.0}
$2p_1$       & $-1$     & 5860+                   \\
$3p_{-1}$    & $0,\pm1$ & 4630\textemdash 5290    \\
$3p_0$       & $0,-1$   & 5800\textemdash 6270    \\
$3p_1$       & $0$      & 4630\textemdash 5270    \\
$3d_0$       & $0,-1$   & 6090\textemdash 6730    \\
$4d_{-2}$    & $0,\pm1$ & 5420\textemdash 6480    \\
$4d_2$       & $0$      & 5420\textemdash 6440    \\
$4f_{-2}$    & $0,\pm1$ & 3650\textemdash 4580    \\
$4f_2$       & $0$      & 3650\textemdash 4560    \\ \hline
\end{tabularx}
\caption{The list of absorption edges that contribute to features in the SDSS J135141 astrophysical linear polarization model, assuming the fiducial magnetic field model. The first column shows the initial hydrogen state labeled by the zero-field quantum numbers $nlm$; the second column labels the transition by $q$, the difference between the initial and final magnetic quantum numbers. The absorption edge features for each respective transition appear in the wavelength range listed in the third column in {\AA}. This wavelength range is equivalent to the range of $\epsilon_{nlmq}$ over the magnetic field strengths present on the surface, $353-705$ MG.
}
\label{tab:AbsEdges}
\end{table}
Note that over the range of magnetic field models and wavelengths analyzed, the axion and astrophysical model contributions to the linear polarization point in the same direction.

$A_{\rm sys}$ is given a zero-mean Gaussian prior distribution in~\eqref{eq:LL}, with variance $\sigma_{\rm sys}^2$.  This prior breaks the degeneracy between the axion signal and the contribution from $A_{\rm sys}$.  We set $\sigma_{\rm sys} = 0.4\%$ since this is the uncertainty quoted in~\cite{2019ASPC..518...93P} on the average linear polarization over this wavelength range and since the uncertainty in~\cite{2019ASPC..518...93P} is systematics dominated.  

We fix $A_{\rm axion}$ and $A_{\rm astro}$ to be positive, since as discussed above these two contributions are polarized in the same direction for this MWD and wavelength range, while $A_{\rm sys}$ is allowed to be both positive and negative.  This means that, for example, the axion and systematic contributions may completely cancel each other, up to the prior contribution from $A_{\rm sys}$.  

We compute the profile likelihood for $A_{\rm sig}$, profiling the likelihood over the nuisance parameters $\{A_{\rm astro},A_{\rm sys},\sigma\}$ for each fixed value of $A_{\rm sig}$.  We then assume Wilks' theorem such that the one-sided 95\% upper limit on $A_{\rm sig}$ is defined through the test statistic $t$ 
\es{eq:t}{
t(A_{\rm axion}) \equiv -2 &\left[  \log p({\bf d}| {\mathcal M}, \{ A_{\rm axion},\hat A_{\rm astro},\hat A_{\rm sys},\hat \sigma\})
\right. \\
&\left. -\log p({\bf d}| {\mathcal M}, \hat {\bm \theta} ) \right] \,,
}
by $t(A_{\rm axion}) \approx 2.71$ for $A_{\rm sig} > \hat A_{\rm sig}$ (see, {\it e.g.},~\cite{Cowan:2010js}).  Here, hatted quantities denote the values that maximize the likelihood.  In the first term in~\eqref{eq:t} the hatted nuisance parameters are those at fixed values of $A_{\rm axion}$.  Performing this analysis on the data illustrated in Fig.~\ref{fig:J135_data} we find $L_{p,{\rm axion}} \lesssim 1.25\%$, where $L_{p,{\rm axion}}$ is the average axion-induced polarization over the wavelength range.  We adopt this upper limit for our analysis. 
Note that the best-fit astrophysical normalization parameter is in fact zero.
In the case where the axion signal has wavelength dependence $L_{p} \propto \lambda^{-2}$, as expected in the large-$m_a$ limit, the limit on $L_{p,{\rm axion}}$  is strengthened to $L_{p,{\rm axion}} \lesssim 0.9\%$.  However, even in the large $m_a$ limit we adopt the upper limit of $1.25\%$ to account for the possibility that the true wavelength dependence of the systematic contribution to the polarization is more complicated than that assumed here.

In Fig.~\ref{fig:J135_data} we illustrate the best-fit model contributions to the data, along with the inferred statistical uncertainty $\sigma$.  The shaded red region shows the allowed values that the axion contribution to $L_p$ could take at 1$\sigma$ significance.  The best-fit model (solid black) has clear evidence of mismodeling; for example, the model systematically under-predicts the data at low $\lambda$ while it over-predicts the data at other wavelengths.  This mismodeling may be from the systematic contribution to the linear polarization having more complicated wavelength dependence than the assumed flat contribution that we take in our analysis.  Still, as the magnitude of the systematic deviations of the best-fit model from the data is smaller, by a factor of a few, than our upper limit on $L_{p,{\rm axion}}$, we hypothesize that a more careful understanding of the instrumental systematic contributions to $L_p$ would be unlikely to significantly affect our estimate of the upper limit. As mentioned previously, the best-fit astrophysical normalization is zero for polarization from bound-free absorption, which we expect to dominate in this wavelength range.  We thus conclude that the observed polarization is likely systematic in nature.  For illustration purposes, we show in Fig.~\ref{fig:J135_data} the linear polarization signal from bound-free emission for the best-fit magnetic field and inclination angle, with an arbitrary normalization. 

A better understanding of the astrophysical background and systematic contributions would be needed to claim evidence for an axion signal.  For this reason we focus in this work only on producing upper limits on $|g_{a\gamma\gamma}|$ and not on looking for evidence for the axion model over the null hypothesis of astrophysical emission only.

\subsubsection{WD radius from {\it Gaia} photometry}

From~\eqref{eq:L_p_dipole_approx} we see that $L_p \propto R_{\rm star}^2$ at low axion masses, so that the limit on $\gagg$ will scale linearly with $R_{\rm star}$.  WDs have radii $\sim$$0.01$$R_\odot$, but as there is scatter from star-to-star it is important to determine the radii on a per-star basis. We infer the WD radius from {\it Gaia} Early Data Release 3 (EDR3) photometry~\cite{2021A&A...649A...3R}. {\it Gaia} has measured SDSS J135141's apparent magnitudes to be $G = 16.4621 \pm 0.0007$, $G_{\rm BP} = 16.486 \pm 0.004$, $G_{\rm RP} = 16.414 \pm 0.005$.

\begin{figure*}[t!]
\includegraphics[width=0.99\textwidth]{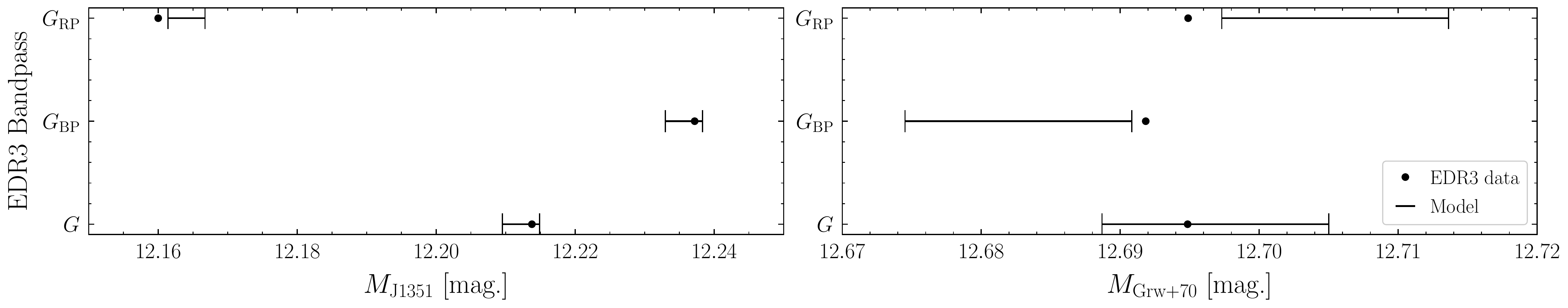}
\caption{\label{fig:rad} (Left) The {\it Gaia} EDR3 data set in the three bandpasses (dots), $G$, $G_{\rm BP}$, and $G_{\rm RP}$, for SDSS J135141. The model from cooling sequences is shown as error bars in each bandpass at the best fit WD mass of $0.7$ M$_\odot$ and age. (Right) The same as the left panel, but now for Grw+70$^\circ$8247 at the best fit WD mass of $1.0$ M$_\odot$.}
\end{figure*}

To infer the WD radius from these data, we use WD cooling sequences~\cite{2020ApJ...901...93B} for WD masses between $0.3$ and $1.2$ M$_\odot$ in steps of $0.1$ M$_\odot$. These sequences provide the expected { EDR3} magnitudes as the WD cools, along with a WD radius. For each mass, we infer the WD radius for SDSS J135141 with a joint Gaussian likelihood over the three bands as a function of age. At a fixed WD mass, we maximize this likelihood over the WD age. To account for possible systematic issues, we additionally maximize over a common uncertainty for $G$, $G_{\rm BP}$, and $G_{\rm RP}$.  That is, we assume that the uncertainties on the magnitudes have a common systematic component, which is added in quadrature with the statistical components and then treated as a nuisance parameter.  We then use the age-radius relation supplied by the cooling sequence to obtain a radius estimate. In the left panel of Fig.~\ref{fig:rad}, we show the {\it Gaia} EDR3 data in each of these bands in absolute magnitudes. We also show the model from the cooling sequence at the best fit WD mass and age.

The best-fit mass for SDSS J135141 is 0.7 M$_\odot$. Within the context of this WD model, the expected radius is $0.0111336 \pm 0.0000003$ R$_\odot$, where the 1$\sigma$ error bars are computed by solving for the ages where the $\Delta\chi^2$ increases by 1 on either side. The WD radius is not highly dependent on age; rather, it is more strongly dependent on mass. Therefore, although the 0.6 and 0.8 M$_\odot$ models are disfavored by the {\it Gaia} data by $\sim$ $4\sigma$, to be conservative we adopt as the radius uncertainties those from assuming the nearby WD masses provided in the cooling sequences. (Ideally, we would use cooling sequences at higher mass resolution than provided in~\cite{2020ApJ...901...93B}.) Using this procedure we infer the radius of SDSS J135141 as $R_{\rm star} = 0.011 \pm 0.001$ R$_\odot$.  Within the uncertainties the most conservative low-mass axion limit is then achieved for $R_{\rm star} = 0.01$ $R_\odot$. 

\subsubsection{Predicted axion-induced polarization signal}

For simplicity we begin by fixing $m_a = 0$ eV and considering how the predicted axion-induced polarization signal varies as a function of the uncertain MWD parameters.  The goal of this exercise is to understand the importance of various sources of modeling uncertainty on the final $g_{a\gamma\gamma}$ upper limit and to determine the most conservative set of fiducial model parameters for computing the upper limit.  In performing these calculations we follow the formalism described in Sec.~\ref{sec:gen_mixing}; specifically, we discretize the surface of the MWD and for each discrete point we solve the mixing equations in~\eqref{eq:mixing-gen} to determine the linear polarization contribution for initially unpolarized rays that leave the surface at that point.  The final polarization signal is the appropriately weighted sum of polarization vectors across the ensemble of all surface points on the hemisphere facing Earth.  We use $10^4$ points on the hemisphere in performing our calculations.

\begin{figure*}[htb]
\begin{center}
\includegraphics[width=0.48\textwidth]{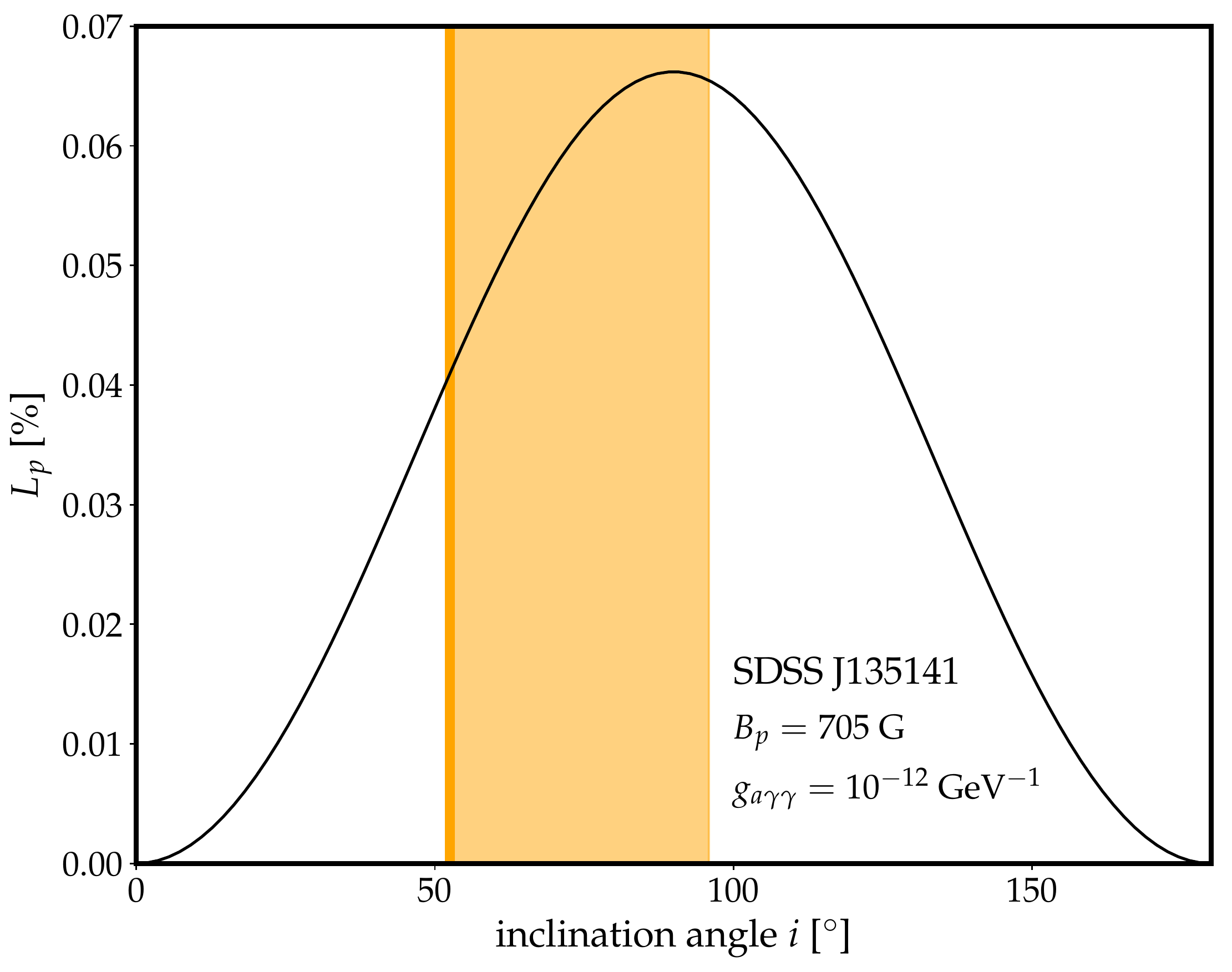}
\includegraphics[width=0.48\textwidth]{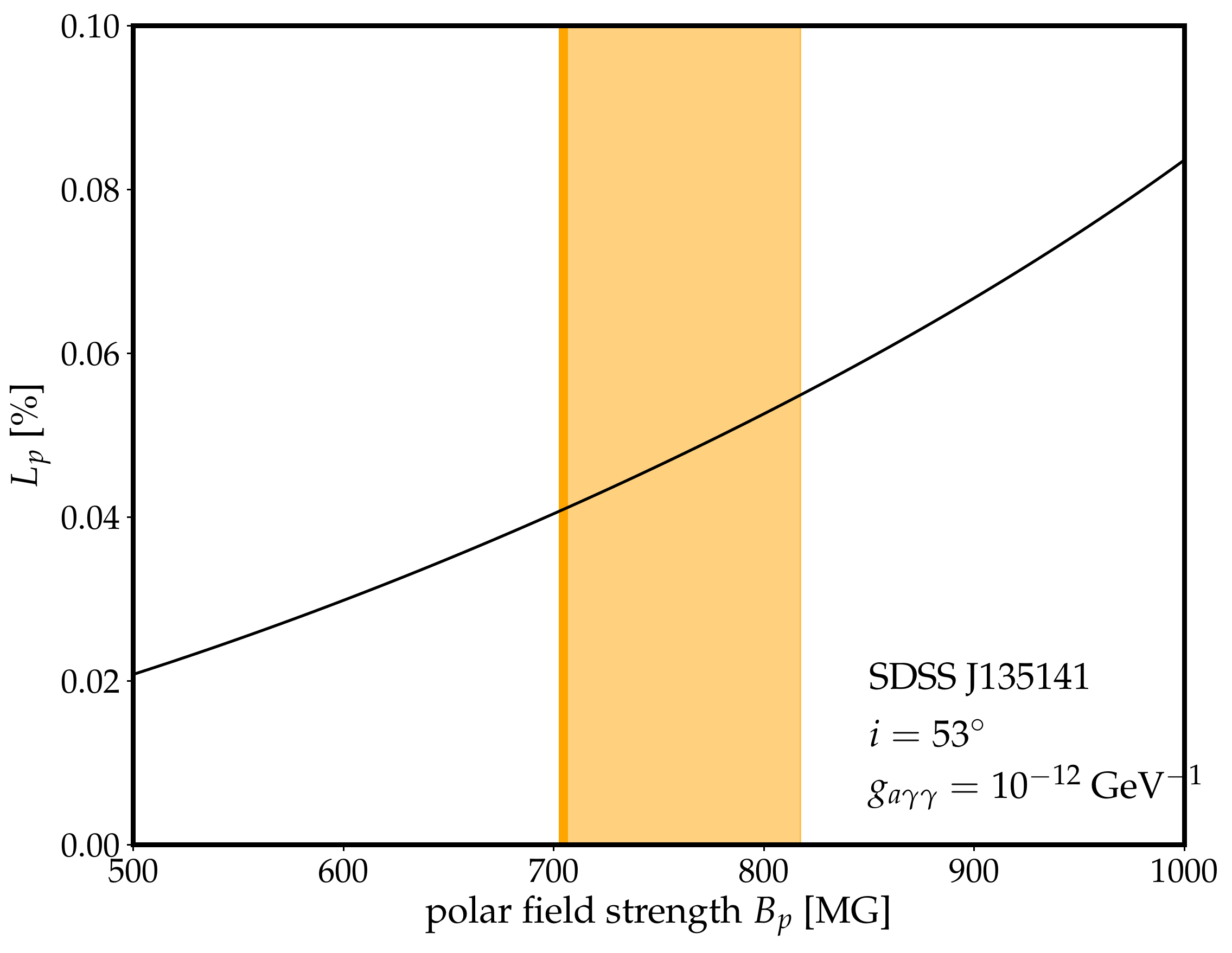}
\caption{\label{fig:angle_J135} 
(Left) The axion-induced linear polarization fraction $L_p$ for SDSS J135141 as a function of the inclination of the magnetic dipole moment relative to the line-of-sight.  The polarization fraction vanishes for $i = 0^\circ$ and $180^\circ$ because in these cases there is no preferred direction for the linear polarization to point.  We highlight in orange the inclination angles preferred at $1\sigma$ by the analysis in~\cite{2009A&A...506.1341K}.  In our fiducial analysis we fix the inclination angle at the value, indicated by vertical orange, within the $1\sigma$ band that leads to the weakest limit.  Note that in the figure we also fix the magnetic field at the lowest value allowed at $1\sigma$, and also the polarization fraction is illustrated for the indicated value of $g_{a\gamma\gamma}$.  Since $L_p \ll 1$, however, the polarization fraction scales approximately quadratically with $g_{a\gamma\gamma}$. (Right) As in the left panel, but illustrating the dependence of $L_p$ on the dipole magnetic field strength.  Note that the inclination angle is fixed at the conservative value indicated in the left panel. The shaded orange region is that preferred at $1\sigma$ by~\cite{2009A&A...506.1341K}; in our fiducial analysis we fix the magnetic field at the value corresponding to the lower edge of this region to be conservative.  In both panels that axion mass is $m_a \ll 10^{-7}$ eV such that $L_p$ is independent of $m_a$.
}
\end{center}
\end{figure*}

In Fig.~\ref{fig:angle_J135} we show how the axion-induced polarization faction from SDSS J135141 varies as functions of the inclination angle $i$ (left panel) and the polar magnetic field strength $B_p$ (right panel).  Note that for this example we fix $R_{\rm star} = 0.01 R_{\rm \odot}$ and $g_{a\gamma\gamma} = 10^{-12}$ GeV$^{-1}$, though since $L_p \ll 1$ the scaling with $g_{a\gamma\gamma}$ is simply $L_p \propto g_{a\gamma\gamma}^2$.  The $L_p$ are computed averaging over the wavelength range   4000~{\AA} to 6500~{\AA} in order to match the polarization data from~\cite{2019ASPC..518...93P}.
The right panel shows, as expected, that increasing field strengths increase the predicted $L_p$; the scaling is roughly quadratic over the range shown.  Shaded in orange is the 1$\sigma$ confidence interval for the polar field strength in the centered dipole model from~\cite{2009A&A...506.1341K}.  The most conservative $B$ field strength in this model is, at $1$$\sigma$, $\sim$705 MG, as indicated by the solid vertical orange line.  The left panel fixes the polar field strength at this value and shows how $L_p$ varies as a function of the inclination angle $i$.  Unsurprisingly, $L_p$ is minimized for $L_p  = 0^\circ$ (or $180^\circ$); the reason, as mentioned previously, is that in these limits for the dipole model there is no preferred direction for the linear polarization to point, so it must vanish.  
Thus, the most conservative value of $i$ at 1$\sigma$ is that closest to zero, which is $i \approx 53^\circ$.

Note that the axion-induced $L_p$ may be approximately a factor of two larger than it is with our fiducial choices, if the $B$-field model parameters are in fact at more fortuitous points in the 1$\sigma$ parameter space.  However, using the most pessimistic allowed magnetic field parameters produces more robust upper limits on $g_{a\gamma\gamma}$. 
It is also important to keep in mind that  
the Zeeman-split lines observed in the spectra give a robust indication of the field strengths on the surface of the MWD on the Earth-facing hemisphere.  The orientation information may be extracted more precisely, however, using circular polarization data, but Ref.~\cite{2009A&A...506.1341K} only used spectral data.  Thus, the orientation determination in the context of the inclination angle measured in Ref.~\cite{2009A&A...506.1341K} is that needed to get the correct distribution of magnetic fields strengths on the Earth-facing hemisphere accounting for the limb darkening.  Analyses of the circular polarization data for this MWD would be useful to better constrain the magnetic field geometry.

In Fig.~\ref{fig:J135_limits} we illustrate the 95\% upper limit on $|g_{a\gamma\gamma}|$ determined from the non-observation of axion-induced polarization from SDSS J135141.  Our fiducial limit is illustrated in solid red and is that obtained with the most pessimistic magnetic field model parameters allowed at 1$\sigma$ from the fits presented in~\cite{2009A&A...506.1341K} ($i \approx 53^\circ$ and $B_{\rm p} = 705$ MG).  In shaded orange we assess  the systematic uncertainty from mismodeling the magnetic field by showing the inferred 95\% limits over the full allowable 1$\sigma$ parameter space for the magnetic field strength and orientation (note that the MWD radius uncertainty is subdominant).  The limit labeled ``best-fit" is that obtained with the best-fit dipole model parameters in~\cite{2009A&A...506.1341K}; the most aggressive limit (labeled optimistic) is found in the offset dipole model by taking the magnetic field at its largest allowed value and $i = 90^\circ$.  

\subsection{Grw+70$^\circ$8247}
\label{sec:GRW}

The MWD Grw+70$^\circ$8247 is thought to have a smaller magnetic field than SDSS J135141, with typical surface field values $\sim$300 MG, but it is an interesting target for axion-induced polarization searches because: (i) modern linear polarization data is available~\cite{2019MNRAS.486.4655B}, and (ii) the magnetic field profile has been well modelled in the context of a harmonic expansion out to $\ell \leq 4$~\cite{2003ASIB..105..175J}.  In particular, Ref.~\cite{2019MNRAS.486.4655B} used the ISIS spectropolarimeter at the William Herschel Telescope to measure the linear polarization of Grw+70$^\circ$8247 in 2015 and 2018.  The linear polarization was measured accross two bands: (i) a blue band (B) from 3700 to 5300 {\AA}, and (ii) a red band (R) from 6100 to 6900 {\AA}.  The linear polarization $L_p$ was found to be non-zero at high significance in the B band, at a level $\sim$3\%, but in the R band the polarization was consistent with zero in both 2015 and 2018. This trend is consistent with that found in earlier observations of $L_p$, going back to 1972~\cite{1972ApJ...171L..11A}, where it is consistently found that the linear polarization is non-zero for wavelengths shorter than $\sim$5000 {\AA} and consistent with zero at lower frequencies. Note that an axion-induced linear polarization signal would be non-zero across the full wavelength range; thus, we may use the R filter data to set a constraint on the possible contribution to the linear polarization from axions.

The R filter linear polarization was measured to be $L_p  = 0.24 \% \pm 0.08 \%$ in 2015 and $L_p = 0.44 \% \pm 0.14 \%$ in 2017~\cite{2019MNRAS.486.4655B}, with uncertainties reflecting photon noise only.  Systematic uncertainties were estimated at $\sim$0.1-0.2\%~\cite{2019MNRAS.486.4655B}.  Assuming the systematic uncertainty is correlated and maximal between the two observing dates, we may combine these results to estimate $L_p = 0.29 \% \pm 0.07_{\rm stat} \% \pm 0.2_{\rm sys} \%$.  Then, we assume Wilks' theorem to estimate $L_p \lesssim 0.29 \% \pm \sqrt{2.71}( 0.07 + 0.2)\% \approx 0.73\%$ at 95\% confidence. Given that the within the R band there is no significant evidence for wavelength dependence~\cite{2019MNRAS.486.4655B}, we use our intuition from the analysis in Sec.~\ref{sec:J135_data} to estimate that the 95\% upper limit on the axion-contribution to $L_p$, accounting for systematic and astrophysical contributions, will be comparable to the estimate above on the total linear polarization limit.  Thus, below we assume $L_{p,{\rm axion}} \lesssim 0.73\%$ at 95\% confidence.

The MWD Grw+70$^\circ$8247 was the first identified MWD~\cite{1970ApJ...161L..77K,1972ApJ...171L..11A} and thus its magnetic field profile is well studied~\cite{1972ApJ...171L..11A,1974ApJ...190L..25L,1975ApJ...196..819L,1985ApJ...292..260A,1992A&A...265..570J,1996ApJ...463..320S,2003ASIB..105..175J,2019MNRAS.486.4655B}.  Additionally, the MWD is known to have a long period, with $P \gtrsim 20$ yrs~\cite{2019MNRAS.486.4655B}.  Ref.~\cite{2003ASIB..105..175J} fit a spherical harmonic magnetic field model including modes with $\ell \leq 4$ to the flux and circular polarization data from Grw+70$^\circ$8247; the result was a field profile of comparable magnitude to the dipole profile but a more non-trivial and twisted spatial distribution.  Interestingly, the dipole and harmonic fits in~\cite{2003ASIB..105..175J} predict nearly identical flux spectra, since the Zeeman effect is only a function of the absolute magnetic field, but the circular polarization prediction from the harmonic model provides a significantly improved fit to the polarization data than the dipole model, since the circular polarization depends on the orientation of the magnetic field.  

The best-fit dipole model from a fit to the flux and circular polarization data for Grw+70$^\circ$8247 was found in~\cite{2003ASIB..105..175J} to have dipole field strength $B_p \approx 347$ MG at an inclination angle $i \approx 56^\circ$.  By contrast, the best-fit harmonic model has $i \approx 75.9^\circ$ and non-trivial $g_\ell^m$ and $h_\ell^m$ through $\ell =4$ that may be found in~\cite{2003ASIB..105..175J}; for example, $g_{10} = 183$ MG, $g_{20} =-40.58$ MG, $g_{30} = 1.39$ MG, and $g_{40} = +1.45$ MG, in the notation of~\eqref{eq:harm}. 

The Grw+70$^\circ$8247 polarization data may naturally be explained by cyclotron absorption. Under the best-fit dipole model, cyclotron absorption will contribute to linear polarization in the range $\sim 3090 - 6170$ {\AA}. This range lies predominantly in the B band.
Thus, we expect the linear polarization to be much larger in the B band than in the R band, as observed in the data. 

 Ref.~\cite{2003ASIB..105..175J} found that in detail the dipole model does not provide a satisfactory fit to the circular polarization data.  
 The harmonic model provided an improved fit to the circular polarization data in~\cite{2003ASIB..105..175J}, though we note that the linear polarization data was not included in their fit. Under the harmonic
 model, the cyclotron absorption contributes to the linear polarization over the full range of both the B and R bands, but the bulk of the support is in the B band (we compute that the mean linear polarization predicted in the B band is $\sim$2 times higher than that in the R band in this model). Therefore, we expect that cyclotron absorption accounts for the fact that higher linear polarization is observed in the B band compared to the R band. On the other hand, note that we do not expect cyclotron absorption to contribute to the linear polarization of the MWD SDSS J135141 in the wavelength range of the data, $4000 - 6500$ {\AA}, because the field is much larger than that of Grw+70$^\circ$8247. For a dipole field strength of 705 MG, as in the most conservative case for SDSS J135141, cyclotron polarization appears only in the wavelength range $\sim 1520 - 3040$ {\AA}. For larger polar field strengths, the cyclotron absorption wavelength range shifts blueward, so that we do not need to consider cyclotron absorption in our analysis of SDSS J135141.

It is interesting to compare the predicted axion-induced polarization signals between the harmonic and dipole models in order to understand the sensitivity of the polarization signal to the magnetic field geometry at the surface of the star.  Note, however, that the photon-to-axion conversion takes place at distances of order multiple $R_{\rm star}$ away from the surface, where the field is dominated by the dipole contribution since the higher-harmonic terms fall off faster with distance from the star. We infer $R_{\rm star}$ for Grw+70$^\circ$8247 in the same way as we do for SDSS J135141, and we obtain $R_{\rm star} = 0.0078 \pm 0.0011$ R$_\odot$ corresponding to $M_{\rm star} = 1.0 \mp 0.1$; to be conservative, we fix $R_{\rm star} = 6.7\times 10^{-3}$ $R_\odot$ throughout this analysis. 
We show the {\it Gaia} data and best-fit cooling sequence model in the right panel of Fig.~\ref{fig:rad}. 

\begin{figure}[t]
\begin{center}
\includegraphics[width=0.48\textwidth]{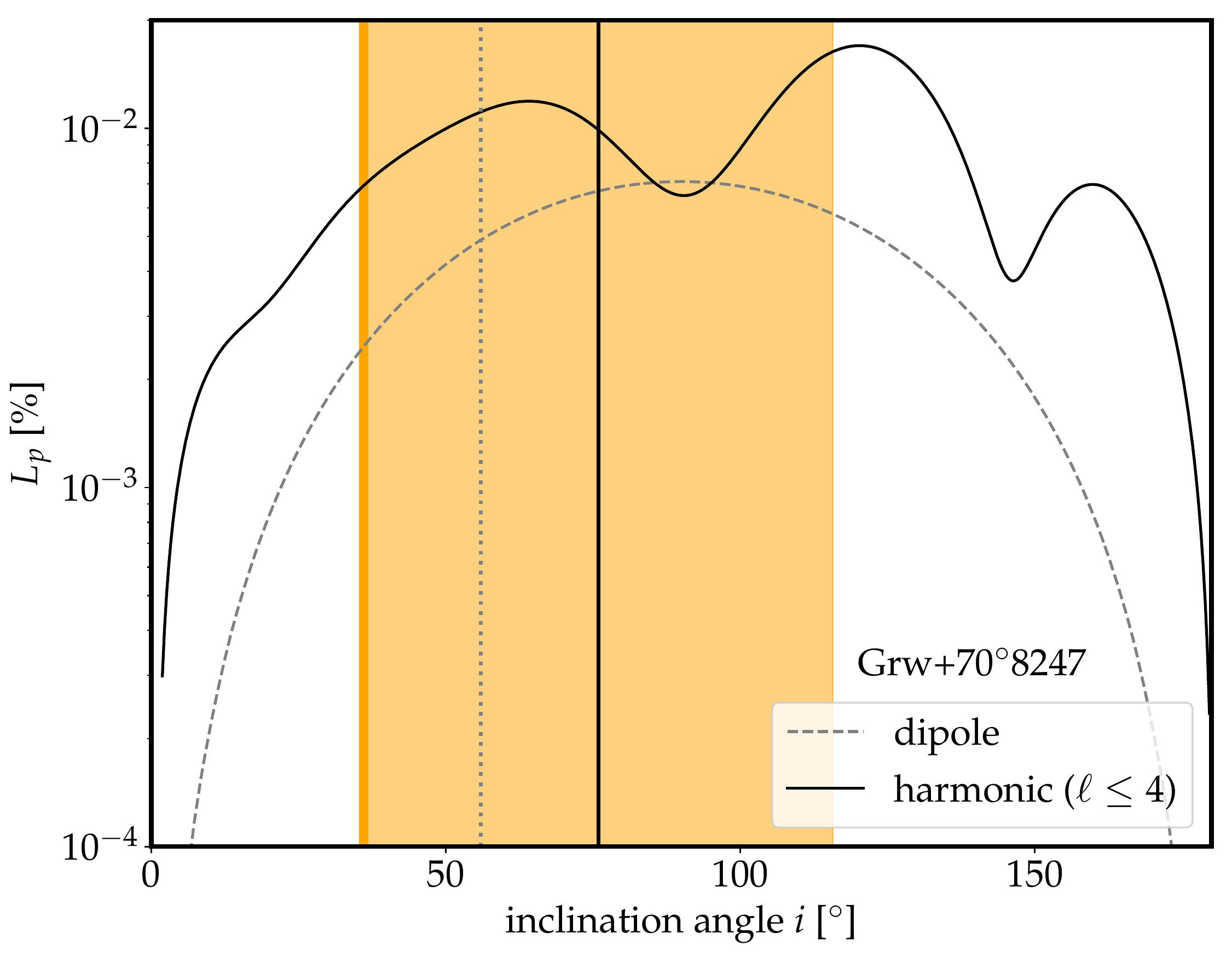}
\caption{\label{fig:GRW_i_plot} 
As in the left panel of Fig.~\ref{fig:angle_J135} but for the MWD Grw$+$70$^\circ$8247.  As in Fig.~\ref{fig:angle_J135} we fix $g_{a\gamma\gamma} = 10^{-12}$ GeV$^{-1}$.  We illustrate the dependence of $L_p$ on the inclination angle for both the dipole fit presented in~\cite{2003ASIB..105..175J}, which has polar field strength $B_p = 347$ MG, and for the best-fit harmonic model (out through $\ell \leq 4$) from ~\cite{2003ASIB..105..175J}.  The best-fit inclination angles for both fits are indicated by the vertical lines (solid for harmonic and dashed for dipole). Note that the harmonic model does not lead to vanishing $L_p$ at $i = 0^\circ$ and $i = 180^\circ$ because their magnetic field profile is not symmetric about the magnetic axis in this case.  Ref.~\cite{2003ASIB..105..175J} does not present uncertainties on their fit parameters, so we estimate that the leading uncertainty arises from the inclination angle. We estimate this uncertainty using the difference between the inclination angles from the dipole and harmonic fits.  In particular, we take the uncertainty on the inclination angle to be twice the difference between the inclination angles measured between the dipole and harmonic fits.  To be conservative we then, in our fiducial analysis, fix the inclination angle in the harmonic model at the indicated value (solid, vertical orange) that leads to the smallest value of $L_p$.  
}
\end{center}
\end{figure}

In Fig.~\ref{fig:GRW_i_plot}
we show the predicted axion-induced linear polarization fraction for $g_{a\gamma\gamma} = 10^{-12}$ GeV$^{-1}$ as a function of the inclination angle $i$, with all other parameters of the dipole and harmonic magnetic field profiles fixed at the best-fit values provided in~\cite{2003ASIB..105..175J}. Note that~\cite{2003ASIB..105..175J} does not provide uncertainties on the inferred model parameters.  As we observe in the previous section when studying SDSS J135141, the dominant uncertainty is likely that arising from the inclination angle.  The best-fit inclination angles quoted in~\cite{2003ASIB..105..175J} are indicated by solid and dashed vertical lines for the harmonic and dipole models, respectively.  We estimate an uncertainty on the harmonic-fit inclination angle $i$ using the difference between the inclination angle measured from the harmonic fit and the dipole fit.  In particular, we take the uncertainty $\sigma_i = 40^\circ$ to be twice the difference between the best-fit inclination angles measured between the two different magnetic field profiles.  Note that this choice of uncertainty is somewhat arbitrary, but it allows us to estimate the possible uncertainty that may arise from mismodeling in the absence of the actual measurement uncertainties.  Additionally, note that in Fig.~\ref{fig:GRW_i_plot} the linear polarization is relatively flat as a function of $i$ for the harmonic fit, except for inclination angles near $0^\circ$ and $180^\circ$ where the dipole and $m =0$ modes do not contribute.  Indeed, it interesting to contrast the harmonic model with the dipole model; the harmonic model generically predicts a larger linear polarization fraction, and the polarization fraction is less sensitive to $i$ in the harmonic case.  The latter point is explained by the fact the dipole model gives rise to vanishing $L_p$ for magnetic axes aligned with the line of sight, while the harmonic model does not because it need not be azimuthally  symmetric about the magnetic axis.  
To be conservative we compute our upper limits on $g_{a\gamma\gamma}$ by fixing $i = 36^\circ$ with the harmonic model, which is the inclination angle over our uncertainty region that gives rise to the lowest $L_p$.

\begin{figure}[htb]
\begin{center}
\includegraphics[width=0.49\textwidth]{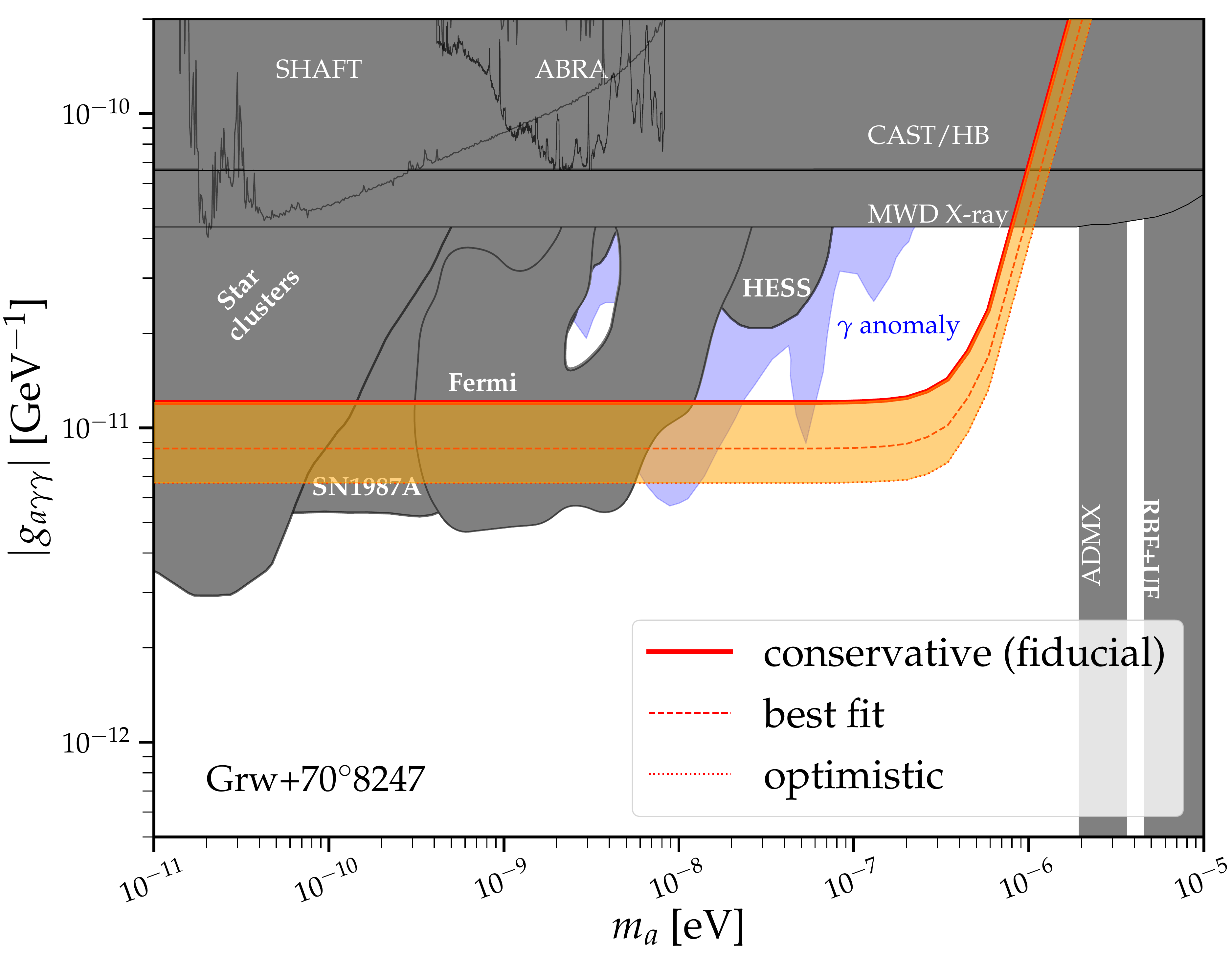}
\caption{\label{fig:GRW_limits} 
As in Fig.~\ref{fig:J135_limits} but for the MWD Grw+70$^\circ$8247.  We compute the upper limit on $g_{a\gamma\gamma}$ using the harmonic magnetic field model.  The orange region arises from varying the inclination angle over the region shown in Fig.~\ref{fig:GRW_i_plot}; the fiducial upper limit is that computed with the inclination angle shown in solid vertical in that figure.  The upper limit computed with the best-fit inclination angle in~\cite{2003ASIB..105..175J} is also indicated.  Note that we fix the MWD radius at $R_{\rm star} = 6.7\times 10^{-3}$ $R_\odot$, which is the smallest value allowed at $1\sigma$ in our analysis, in order to be conservative.    
}
\end{center}
\end{figure}

In Fig.~\ref{fig:GRW_limits}
we illustrate the 95\% upper limit on $g_{a\gamma\gamma}$ as a function of the axion mass $m_a$, as in Fig.~\ref{fig:J135_limits}, for the Grw+70$^\circ$8247 analysis.  We compute the 95\% upper limit under three assumptions: (i) the harmonic model with $i = 36^\circ$, which is our fiducial limit; (ii) the harmonic model at the best-fit $i \approx 75.9^\circ$, and (ii) the harmonic model with $i \approx 116^\circ$, which is the inclination angle within our 1$\sigma$ band that gives rise to the maximal $L_p$ prediction.    
The shaded band in Fig.~\ref{fig:J135_limits} covers this range of possibilities and is an estimate of the systematic uncertainty from magnetic field mismodeling.

\subsection{Additional MWDs}
\label{sec:other}

In this section we comment on additional promising MWDs where linear polarization data is already available or where acquiring polarization data should be a priority for the future.  First note that Ref.~\cite{Gill_2011} suggests upper limits on $g_{a\gamma\gamma}$ at the level of $|g_{a\gamma\gamma}| \lesssim (5 - 9) \times 10^{-13}$ GeV$^{-1}$ using the linear polarization data from the MWDs PG 1031+234 and SDSS J234605+38533.  We begin by revisiting these MWDs to assess the robustness of the upper limits from these stars.  

A fit of the centered dipole magnetic field model to the intensity spectra for the MWD SDSS J234605+38533 measured by the SDSS resulted in a polar field strength $B_p = 798 \pm 164$ and inclination angle $i = 2.5^\circ \pm 1.1^\circ$~\cite{2009A&A...506.1341K}.  Note, however, that this analysis only consider intensity spectra and not circular polarization, and so the orientation angle is only constrained by producing the correct distribution of surface field strengths not directly by the orientation of the magnetic field structure.  Indeed, in the context of the offset dipole model a comparable magnetic field strength was found but for $i  = 87^\circ \pm 15^\circ$~\cite{2009A&A...506.1341K}.  Ref.~\cite{SDSS:2005uus} measured a linear polarization from SDSS J234605+38533 of $L_p \approx 1.33\%$, though with no uncertainties quoted, across the wavelength range $4200$ {\AA} to $8400$ {\AA} using the SPOL instrument on the Steward Observatory  Bok Telescope and the Multiple Mirror Telescope (MMT) on Mt. Hopkins (see~\cite{SDSS:2003vmj} for details).  Without uncertainties on the $L_p$ measurement, it is difficult to estimate the 95\% upper limit on the linear polarization.  For concreteness, let us imagine that the upper limit is $L_p \lesssim 2\%$ over this wavelength range.  To set a conservative upper limit, we take $i = 1.4^\circ$ for the centered dipole with $B_p = 634$ MG, since this is the most conservative scenario consistent within the 1$\sigma$ uncertainties for $B_p$ and $i$.  We also fix $R_{\rm star} = 0.01 \, \, R_\odot$ for definiteness.  For $m_a \ll 10^{-6}$ eV we find that this then translates into a limit $|g_{a\gamma\gamma}| \lesssim 2.1 \times 10^{-11}$ GeV$^{-1}$, though it is important to remember that this is an estimate since no rigorous upper limit on $L_p$ is available.  This upper limit is comparable to the conservative upper limit from Grw+70$^\circ$8247, weaker than the conservative upper limit from SDSS J135141, and significantly weaker than the $|g_{a\gamma\gamma}| \lesssim (5 - 9) \times 10^{-13}$ GeV$^{-1}$ upper limit quoted from this MWD and PG 1031+234 in~\cite{Gill_2011}.  However, it is possible that the limit from SDSS J234605+38533 could be improved with a better determination of the magnetic field geometry, since {\it e.g.} the off-set dipole model prefers much larger inclination angles. 

\begin{figure*}[htb]
\begin{center}
\includegraphics[width=0.48\textwidth]{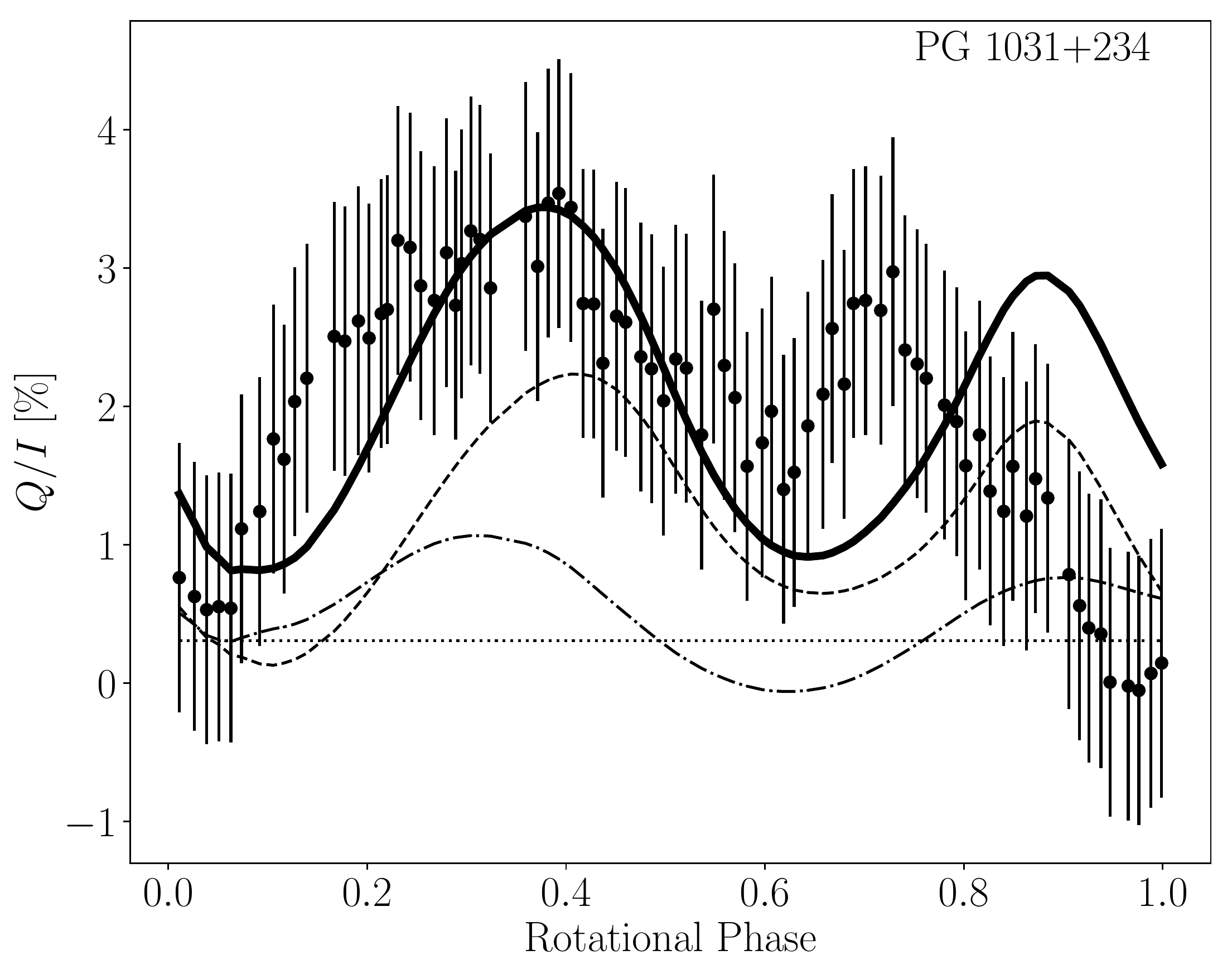}
\includegraphics[width=0.48\textwidth]{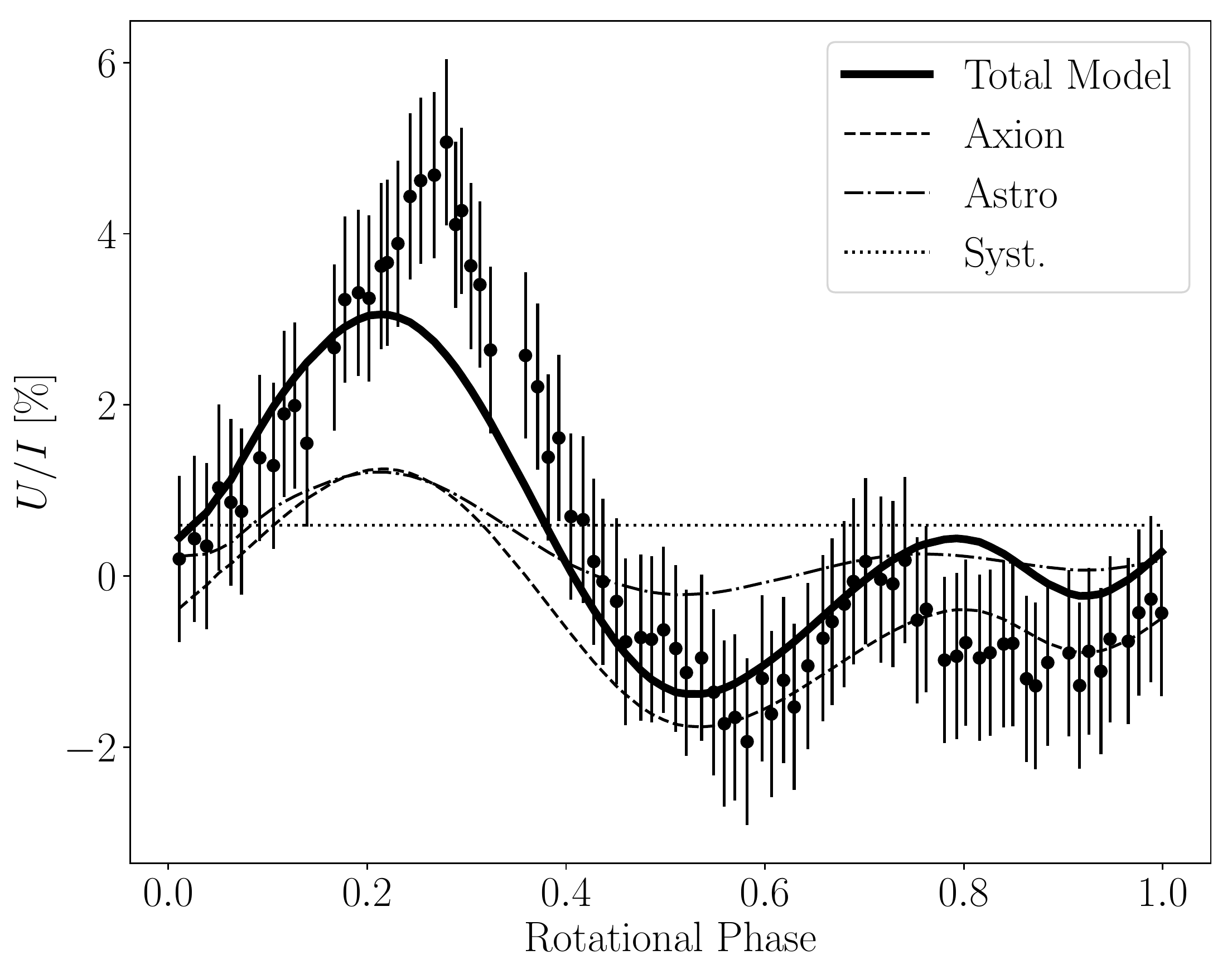}
\caption{\label{fig:PG_1031} 
The linear polarization data from~\cite{1986ApJ...309..218S} for PG 1031+234 presented as ratios of the Stokes parameters $Q$ (left) and $U$ (right) relative to the intensity $I$.  We fit a model consisting of an axion, astrophysical, and systematic contributions to the joint $Q/I$ and $U/I$ data, treating the statistical uncertainty as a nuisance parameter.  We display the best-fit joint model, in addition to the best-fit components.  The uncertainties on the data points are the best-fit uncertainties from maximum likelihood estimation of the associated hyperparameter.  The magnetic field model consists of two dipoles, with one being offset, and thus the axion and astrophysical contributions have varying phase differences over the rotational phase of the MWD.  We estimate the constraint $|g_{a\gamma\gamma}| \lesssim 8.8 \times 10^{-12}$ GeV$^{-1}$ at 95\% confidence for $m_a \ll 10^{-7}$ eV, subject to the caveat that the magnetic field model is fixed at the best-fit model from~\cite{1986ApJ...309..218S}.  The best-fit axion coupling, corresponding to the illustrated curve, is $g_{a\gamma\gamma} \approx 7.4\times10^{-12}$ GeV$^{-1}$.  
}
\end{center}
\end{figure*}

Next, we consider  PG 1031+234, which was the second MWD from~\cite{Gill_2011} that led to the proposed upper limit $|g_{a\gamma\gamma}| \lesssim (5 - 9) \times 10^{-13}$ GeV$^{-1}$ for low axion masses. This MWD is unique relative to the MWDs considered so far in this work in that it has a period $\sim$3 hr 24 min that leads to observable oscillations in the polarization and flux spectra~\cite{1986ApJ...309..218S,1992A&A...259..143P}.  The linear polarization data from~\cite{1986ApJ...309..218S} stacked over the rotational phase of the MWD in the band 3200--8600 {\AA} is illustrated in Fig.~\ref{fig:PG_1031}; the left (right) panel shows the Stokes parameter ratio $Q/I$ ($U/I$).  These ratios are inferred from the data in~\cite{1986ApJ...309..218S} using the linear polarization data and the polarization angle.  The uncertainties in Fig.~\ref{fig:PG_1031} are estimated during the model fitting process, as described shortly.

The MWD  PG 1031+234 was modeled in~\cite{1986ApJ...309..218S} as having a centered dipole field with a polar field strength $\sim$500 MG and a small magnetic hot-spot that has a much larger field strength $\sim$$10^3$ MG.  More specifically, Ref.~\cite{1986ApJ...309..218S} showed that the following magnetic field model was able to explain the major features observed in the flux, circular polarization, and linear polarization data by using radiative transfer models to estimate to the polarization and absorption signals at different points on the MWD surface.  Their model included a centered dipole with polar field strength $B_c \approx 400$ MG and magnetic axis inclined by 35$^\circ$ relative to the rotation axis. The rotation axis is at an inclination angle of $i = 60^\circ$ relative to the line of sight.  The magnetic hot-spot is modeled by an offset dipole with magnetic axis inclined at 55$^\circ$ relative to the rotation axis, polar surface field strength of $10^3$ G, and offset $z_{\rm off}  = 0.4 \, \, R_{\rm star}$ along the magnetic axis.  The magnetic hot-spot precedes the centered dipole by a phase of 120$^\circ$.  In Fig.~\ref{fig:PG_1031} we adjust the phase such that zero corresponds to the transit of the centered dipole. The radiative transfer calculation in~\cite{1986ApJ...309..218S} using this model was able to explain the broad features observed in both the circular and linear polarization data, though an axion signal would only contribute to the linear polarization.

We compute the astrophysical contribution to the linear polarization using a similar method to that in~\cite{1986ApJ...309..218S}. In particular, we use the formalism in~\cite{lamb_sutherland_1974}, including both the bound-free and cyclotron contributions to the polarization, as cyclotron absorption is expected to contribute in the wavelength band of the observations. We compute the astrophysical Stokes parameters averaged over wavelengths and over $\sim 10^5$ points on the observable hemisphere at a fixed phase. We repeat this process over all of the rotational phases of the MWD. 
Note that we assign the astrophysical model two unconstrained nuisance parameters that independently normalize the amplitudes of the linear polarization contributions from bound-free and cyclotron absorption.  

We compute the axion-induced linear polarization signal for the magnetic field model described above assuming $m_a \ll 10^{-7}$ eV. The polarization signal is illustrated in Fig.~\ref{fig:PG_1031} for the best-fit coupling $g_{a\gamma\gamma} \approx 7.4 \times 10^{-12}$ GeV. 

In addition to the astrophysical and axion contributions to the polarization, we separately add in phase-independent systematic contributions to $Q/I$ and $U/I$.  These contributions are to allow for instrumental effects that could bias $Q/I$ or $U/I$ away from zero.  We then construct a joint likelihood over the $Q/I$ and $U/I$ data, with the axion and astrophysical models contributing to both ratios.  Since we do not know the alignment of the MWD on the sky,  we allow for an additional nuisance parameter that rotates the projection of the MWD on the sky.  Note, however, that the astrophysical and axion contributions rotate by the same amount for a given orientation.  Lastly, we determine the uncertainties on the data in a data-driven way by assigning the uncertainties to be hyperparameter that is treated as a nuisance parameter and determined by maximum likelihood estimation, as in {\it e.g.}~\eqref{eq:LL}.  In total, we thus have our signal parameter $g_{a\gamma\gamma}$ and six additional nuisance parameters.  

The best fit of the joint signal and background model is illustrated in Fig.~\ref{fig:PG_1031}, along with the best-fit component contributions.  
Note that while the model is able to describe the broad features in the data,   there is clear evidence for mismodeling across the phase of the MWD.  On the other hand, our goal here is not to derive a precise limit, since for example we do not account for uncertainties on the magnetic field model, but rather to illustrate key points behind the phase-resolved analysis and to roughly estimate the magnitude of the limit that may emerge from a more careful analysis.

Importantly, the $Q/I$ and $U/I$ axion and astrophysical contributions vary independently over the phase of the MWD, since they depend differently on the observable magnetic field geometry.  Thus, large cancellations between the axion and astrophysical contributions are not possible across all phases and for both $Q/I$ and $U/I$.  This leads to the result that the 95\% upper limit on $g_{a\gamma\gamma}$, as determined from the profile likelihood, is estimated as $|g_{a\gamma\gamma}| \lesssim 8.8 \times 10^{-12}$ GeV$^{-1}$, which is relatively close to the best-fit axion coupling of $g_{a\gamma\gamma} \approx 7.4 \times 10^{-12}$ GeV$^{-1}$.  We caution, however, that this upper limit should be treated with caution, since it does not account for uncertainties on the magnetic field profile and since the fits in Fig.~\ref{fig:PG_1031} show evidence for mismodeling.  Still, it is striking that our estimate for the upper limit around an order of magnitude weaker than the upper limit estimate in~\cite{Gill_2011} for the same MWD. 

The example of PG 1031+234 highlights how rotational-phase resolved data may be useful in the context of the axion-induced linear polarization search.  This example motivates, in particular, a search for axion-induced polarization from the MWD RE J0317-853.  This MWD is rotating quickly with a period $\sim$725 s~\cite{1995MNRAS.277..971B}. The magnetic field varies across the surface over the rotation period between $\sim$200 -- 800 MG~\cite{Burleigh:1998pqa}.  Moreover, Ref.~\cite{Burleigh:1998pqa} presented a model for the magnetic field structure in terms of a harmonic expansion through $\ell \leq 3$ with a magnetic axis offset from the rotation axis, which is at a non-zero angle to the line-of-sight.  Unfortunately, no linear polarization data is available for RE J0317-853 at present, but acquiring such data and interpreting it in the context of the axion model should be a priority.  We note that~\cite{Dessert:2021bkv} recently used $X$-ray data from RE J0317-853 to search for axion-induced hard $X$-ray signals.
\begin{table}[]
\centering
\setlength{\tabcolsep}{7pt}
\begin{tabularx}{0.37\textwidth}{cc}
\hline
MWD Name                    & $B_p$ [MG]        \\ \hline
RE J0317-853                & $\sim$$200-800$      \\
SDSS J033320.36+000720.6    & $849 \pm 42$      \\
SDSS J002129.00+150223.7    & $531 \pm 64$      \\
SDSS J100356.32+053825.6    & $672 \pm 119$     \\
HE 1043-0502                & $\sim$$820$        \\
SDSS J120609.80+081323.7    & $761 \pm 282$     \\
ZTF J190132.9+145808.7      & $\sim$$600-900$    \\ \hline
\end{tabularx}
\caption{MWDs without existing linear polarization data but which would be promising targets for future axion searches, due to their large magnetic fields. The magnetic fields for these targets were determined by Refs.~\cite{2009A&A...506.1341K,2001MNRAS.328..203S,Caiazzo:2021xkk}.
}
\label{tab:MWDs}
\end{table}
A list of MWDs which do not currently have linear polarization data but with large magnetic fields, including RE J0317-853, is in Tab.~\ref{tab:MWDs}. In addition to high-resolution linear polarization data from the MWDs, circular polarization data would be useful in order to better constrain the magnetic geometries of these MWDs using radiative transfer theory.

\section{Discussion}

In this work we model how axions may induce polarization signals in the otherwise unpolarized thermal emission from MWD surfaces.  We show that MWDs are optimal targets for axion-induced polarization searches because they have large magnetic fields but not so large that the Euler-Heisenberg Lagrangian suppresses the photon-to-axion conversion probability.  Larger stars with lower magnetic field strengths have reduced conversion probabilities because of the axion-to-photon mixing term, while the more compact NSs, which have stronger magnetic fields, are in the regime where the Euler-Heisenberg term suppresses the mixing by modifying the photon dispersion relation relative to that of the axion.  At the same time, the predicted astrophysical backgrounds to the linear polarization from MWDs are minimal, relative to {\it e.g.} those from NSs, and induced by polarization-dependent radiative transfer processes for initially unpolarized surface emission propagating through the thin, magnetized MWD atmospheres.

The axion-induced polarization signal from MWDs was previously discussed in~\cite{Lai:2006af,Gill_2011}, where it was claimed that linear polarization data from the MWDs SDSS J234605+38533 and PG 1031+234 may already constrain the axion-photon coupling to $|g_{a\gamma\gamma}| \lesssim (5-9) \times 10^{-13}$ GeV$^{-1}$ for low axion masses $m_a \ll 10^{-7}$ eV.  We provide a simple formalism for predicting the axion-induced polarization signal, which only involves the field configuration far away from the MWD surface, and we show that these previous limits are likely overstated.  However, we present analyses from two MWDs with dedicated linear polarization data and well-measured magnetic field distributions: SDSS J135141 and GRW$+$70$^\circ$8247.  The conservative upper limit from SDSS J135141, which is $|g_{a\gamma\gamma}| \lesssim 5.4 \times 10^{-12}$ GeV$^{-1}$, is the strongest to-date over a large region of axion masses and strongly disfavors the axion interpretation of the previously-observed gamma-ray transparency anomalies.  Future linear polarization measurements, in conjunction with dedicated modeling efforts for the magnetic field geometries and astrophysical linear polarization backgrounds, towards promising targets such as RE J0317-853 could further strengthen these limits and perhaps unveil evidence for low-mass axions.     

\section*{Acknowledgments}

We thank J. Foster and Georg Raffelt for useful discussions.  C.D. and B.R.S. were supported  in  part  by  the  DOE  Early Career  Grant  DESC0019225. This research used resources from the Lawrencium computational cluster provided by the IT Division at the Lawrence Berkeley National Laboratory, supported by the Director, Office of Science, and Office of Basic Energy Sciences, of the U.S. Department of Energy under Contract No.  DE-AC02-05CH11231.


\bibliography{bibliography}

\end{document}